\documentclass[aps,pre]{revtex4-1}

\usepackage{graphicx}% Include figure files
\usepackage{dcolumn}% Align table columns on decimal point
\usepackage{bm}% bold math
\usepackage[utf8]{inputenc}
\usepackage{mathtools}
\usepackage[T1]{fontenc}
\usepackage{mathptmx}
\usepackage[utf8]{inputenc}
\usepackage{graphicx}
\usepackage{subcaption}
\usepackage{amsmath}
\usepackage{amssymb}
\usepackage{dcolumn}% Align table columns on decimal point
\usepackage{bm}% bold math
\usepackage{xr}
\usepackage{cleveref}

\begin{document}

\title{Kinetic mechanisms of crumpled globule formation}
\author{Artem Petrov}
\email{petrov.ai15@physics.msu.ru}
	\affiliation{Faculty of Physics, Lomonosov Moscow State University, 119991 Moscow, Russia}
	\author{Pavel Kos}
	\affiliation{Faculty of Physics, Lomonosov Moscow State University, 119991 Moscow, Russia}
	\affiliation{Semenov Institute of Chemical Physics, 119991 Moscow, Russia}
	\author{Alexander Chertovich}
	\affiliation{Semenov Institute of Chemical Physics, 119991 Moscow, Russia}
	\affiliation{Faculty of Physics, Lomonosov Moscow State University, 119991 Moscow, Russia}
	\date{\today}

\begin{abstract}
Homopolymer chain with beads forming pairwise reversible bonds is a well-known model in polymer physics. We studied kinetics of homopolymer chain collapse, which was induced by pairwise reversible bonds formation. We compared kinetic mechanism of this coil-globule transition with the mechanism of collapse in a poor solvent. We discovered, that coil-globule transition occurs sufficiently more homogeneously on different scales, if collapse is induced by pairwise reversible bonds formation. This effect leads to formation of transient structures, which are not similar to the classical pearl-necklace conformations formed during collapse in a poor solvent. However, both types of collapse lead to formation of a metastable state of crumpled globule, which is one of the well-known models of interphase chromatin structure in different organisms. Moreover, we found out that stability and dynamics of this state can be controlled by fraction of reversible bonds and bond lifetime.
\end{abstract}

\maketitle

\section{Introduction}
Kinetics of coil-globule transition in a homopolymer chain is one of the unsolved problems in the modern polymer physics. This issue is closely related to the problem of protein folding \cite{rostiashvili2003collapse,lappala2013raindrop}, as well as chromatin structure description. There is experimental evidence of two-stage process of coil-globule transition in a poor solvent \cite{chu1995two,xu2006first,ye2007many}. Moreover, various scaling and self-consistent field theories describe the process of homopolymer chain collapse in a poor solvent \cite{abrams2002collapse,kikuchi2005kinetics,rostiashvili2003collapse,kuznetsov1996kinetic,halperin2000early}. There are different computer simulation studies on that topic as well \cite{abrams2002collapse,lappala2013raindrop,schram2013stability,dynamicsofvolumecollapse,kikuchi2005kinetics,bunin2015coalescence}. The aforementioned research suggests that a homopolymer chain forms blobs on a local scale (i.e. pearl-necklace structure \cite{halperin2000early}) on the early stages of collapse. Blobs grow in size continuously, merging together, and the chain collapses globally. However, size distribution of blobs during collapse under active compression can be very nontrivial \cite{bunin2015coalescence}. It is also unclear how chain structure changes in time inside blobs during collapse, and theoretical works do not describe this evolution. It was suggested, that the partially mixed crumpled globule is stabilized on a large scale by topological constraints \cite{grosberg1988role,dynamicsofvolumecollapse,bunin2015coalescence} after global collapse of the chain. Regulation of stability of the crumpled globule steady-state is also an important unsolved problem \cite{dynamicsofvolumecollapse,schram2013stability}. Cross-links may stabilize crumpled globule state \cite{schram2013stability}.

Crumpled globule is a well-known model of chromatin packing in cell nucleus. Spatial chromatin structure can be characterized by the probability of finding two loci spatially close to each other, depending on the genomic distance between the loci, $P(s)$. This method became popular after Hi-C technique invention \cite{Hi-C}. It was revealed that chromatin is packed hierarchically: on a scale of several thousand base pairs domain-like structures (topologically associating domains, or TADs) are formed \cite{tadspositions}. In most species on a larger scale TADs form compartments, which segregate into active and inactive parts \cite{tadspositions, grosbergreview, drosophilaconf}. Moreover, $P(s)$ dependencies appeared to be significantly different for various species and cell types \cite{mouseconf, yeastconf, drosophilaconf, Hi-C, barbieri2012complexity} (Fig. S11). This may indicate different spatial organization of chromatin. There is evidence of crumpled globule model applicability for chromatin structure description on submegabase scale \cite{Hi-C, tamm2015anomalous, tadspositions, fractalCDhumanmodel}, as well as subdiffusion exponent prediction in various organisms \cite{telomeres1, kepten2013improved,weber2012nonthermal}. However, it is a challenge to obtain a long-living crumpled globule state: collapse of a chain, induced by the artificial gravitation-like potential \cite{mirny2011fractal} or collapse in a very poor solvent \cite{dynamicsofvolumecollapse} seems unrealistic. Fractal globule state observed in  Strings and Binders Switch (SBS) model \cite{barbieri2012complexity} exists in a narrow range of binder concentrations, and therefore it is unstable to fluctuations. There were attempts to describe large-scale chromatin structure by the block copolymer models with different types of interactions \cite{chertovichcopolymermodeldrosophila, jost2014modeling, barbieri2012complexity}. However, dynamical behavior of the aforementioned models remains obscure. There is a distinct class of active matter-based models of chromatin organization as well \cite{ganai2014chromosome,vandebroek2015dynamics,osmanovic2017dynamics}. These models explain several non-trivial dynamical effects, presumably connected with active noise \cite{weber2012nonthermal}. However, there is experimental evidence that active noise affects only apparent diffusion coefficient of loci motion, but subdiffusion exponent remains unchanged \cite{weber2012nonthermal}. Active matter-based models are focused mostly on description of dynamical behavior of polymers in active medium, but conformational properties of the model chains are not studied in detail. Therefore, the problem of large-scale chromatin structure description remains unsolved.

\begin{figure}[htbp]
	\includegraphics[width=0.99\textwidth]{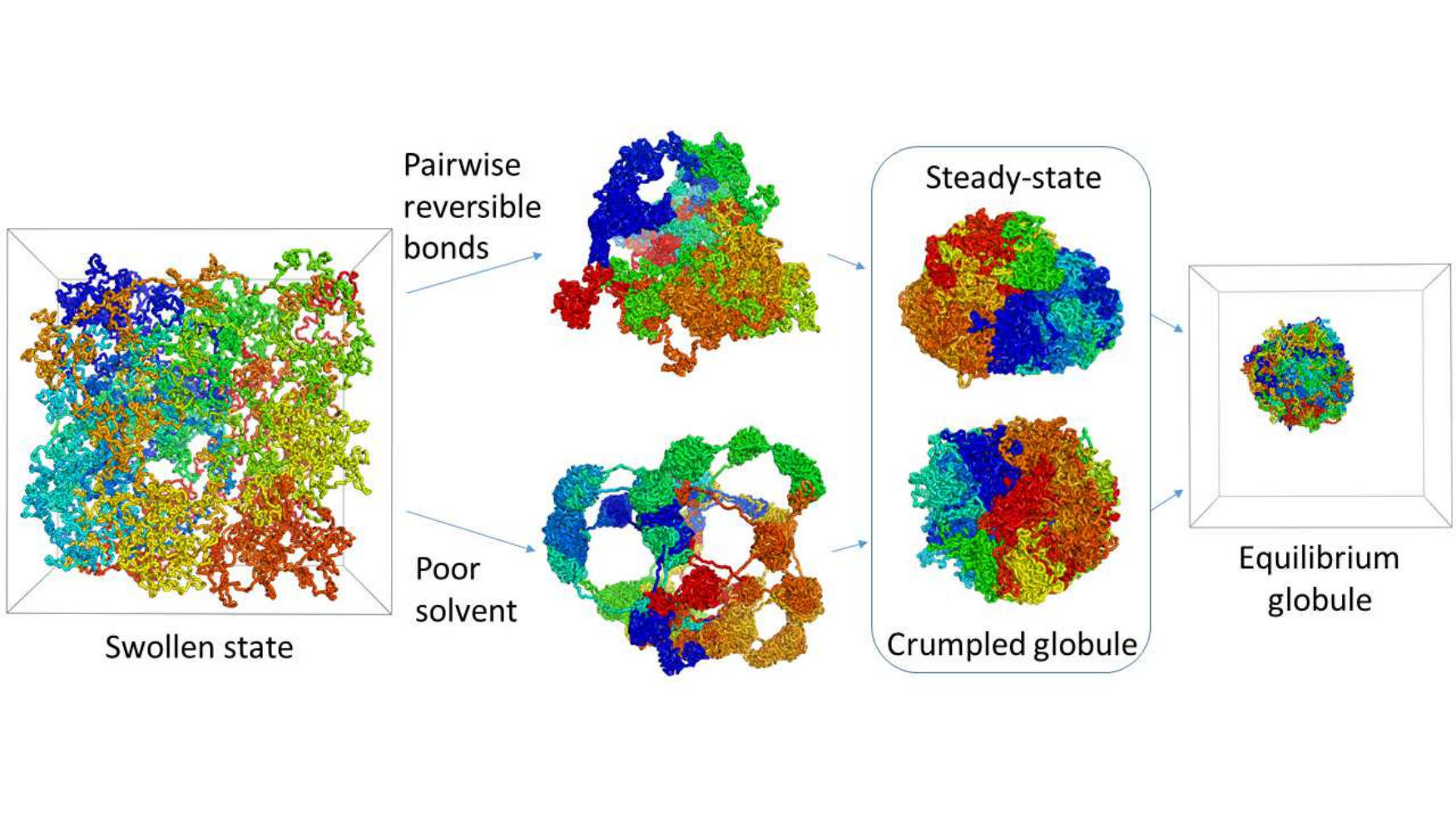}
    \caption{General overview of the systems under study. Both System with Pairwise Bonds (SPB) and System in a Poor Solvent (SPS) collapse to a similar steady crumpled globule state, but the intermediate conformations are considerably different.}
    \label{general}
\end{figure}

In this work, we describe a method, which allows to compare kinetic mechanisms of collapse induced by different types of intrachain interactions. We applied this method to compare collapse mechanisms from an initially swollen and spatially confined chain, see Fig. \ref{general}. Two types of collapse have been studied: collapse induced by pairwise reversible bonds formation in athermal solvent and collapse in a poor solvent. Coil-globule transition induced by pairwise reversible bonds formation was first described by I.M. Lifshitz et al. \cite{Lifshits1976}. However, kinetics of collapse, as well as possible metastable transient structures, have not been described in the literature yet. We have discovered differences in kinetic mechanisms of collapse in poor solvent and coil-globule transition induced by pairwise reversible bonds formation. Strikingly, we have observed formation of a steady-state similar to the crumpled globule in both cases. We have also described effect of pairwise reversible bonds presence on dynamics inside crumpled globule and equilibrium globule to assess stability of the steady-state.
Based on our results, we propose a novel simple model of large-scale chromatin organization.

\section{Methods}

We used Dissipative Particle Dynamics (DPD) method to perform simulations. DPD is a mesoscale molecular dynamics with explicit solvent and soft potential for the conservative force. One of the most important features of DPD method is that it conserves hydrodynamic interactions, which may affect dynamics of collapse in a poor solvent \cite{kuznetsov1996kinetic,kikuchi2005kinetics}. Moreover, DPD conserves momentum, which can also affect dynamics of a polymer chain \cite{pastorino2007comparison}. We do not describe this method in detail, more information about DPD can be found in Ref. \cite{groot1997dissipative}. We only note here, that we added one more conservative force, describing chain connectivity of beads: $\overrightarrow{F}_{ij}^b = -K(r_{ij}-r_0)\hat{\boldsymbol{r}_{ij}}$, here $K$ is the spring constant (bond stiffness) and $r_{ij}$ is the distance between i-th and j-th bead in the chain, $r_0$ is the equilibrium bond length, and $\hat{\boldsymbol{r}_{ij}} = \overrightarrow{r}_{ij}/r_{ij}$ ($\overrightarrow{r}_{ij}$ is the vector, pointed from the center of i-th bead to the center of j-th bead). We emphasize that there is no angle potential in the chain.

\begin{figure}[htbp]
    \centering
    \begin{subfigure}{0.45\textwidth}
	\includegraphics[width=\linewidth,height=\textheight,keepaspectratio]{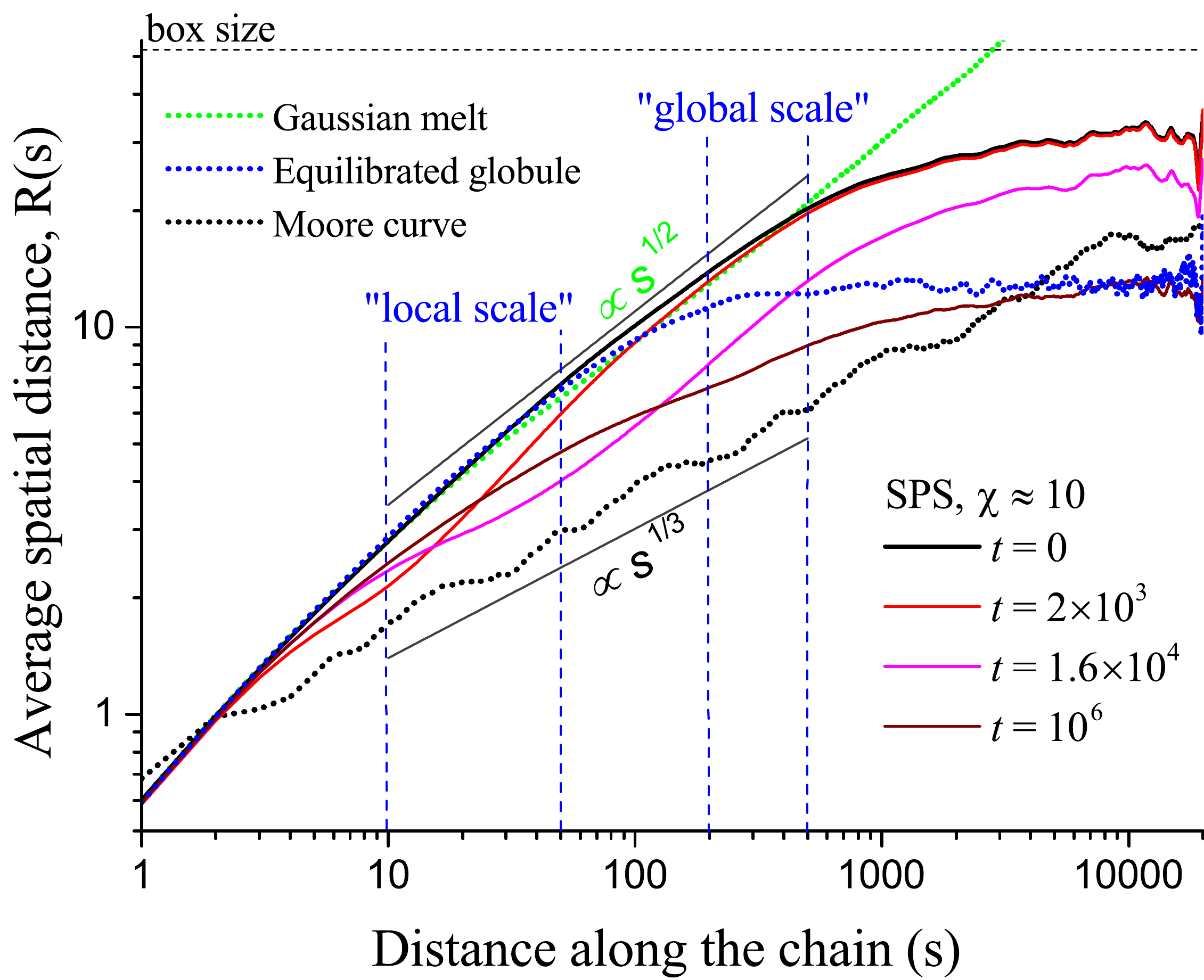}
	\caption{}
	\end{subfigure}
	\begin{subfigure}{0.45\textwidth}
	\includegraphics[width=\linewidth,height=\textheight,keepaspectratio]{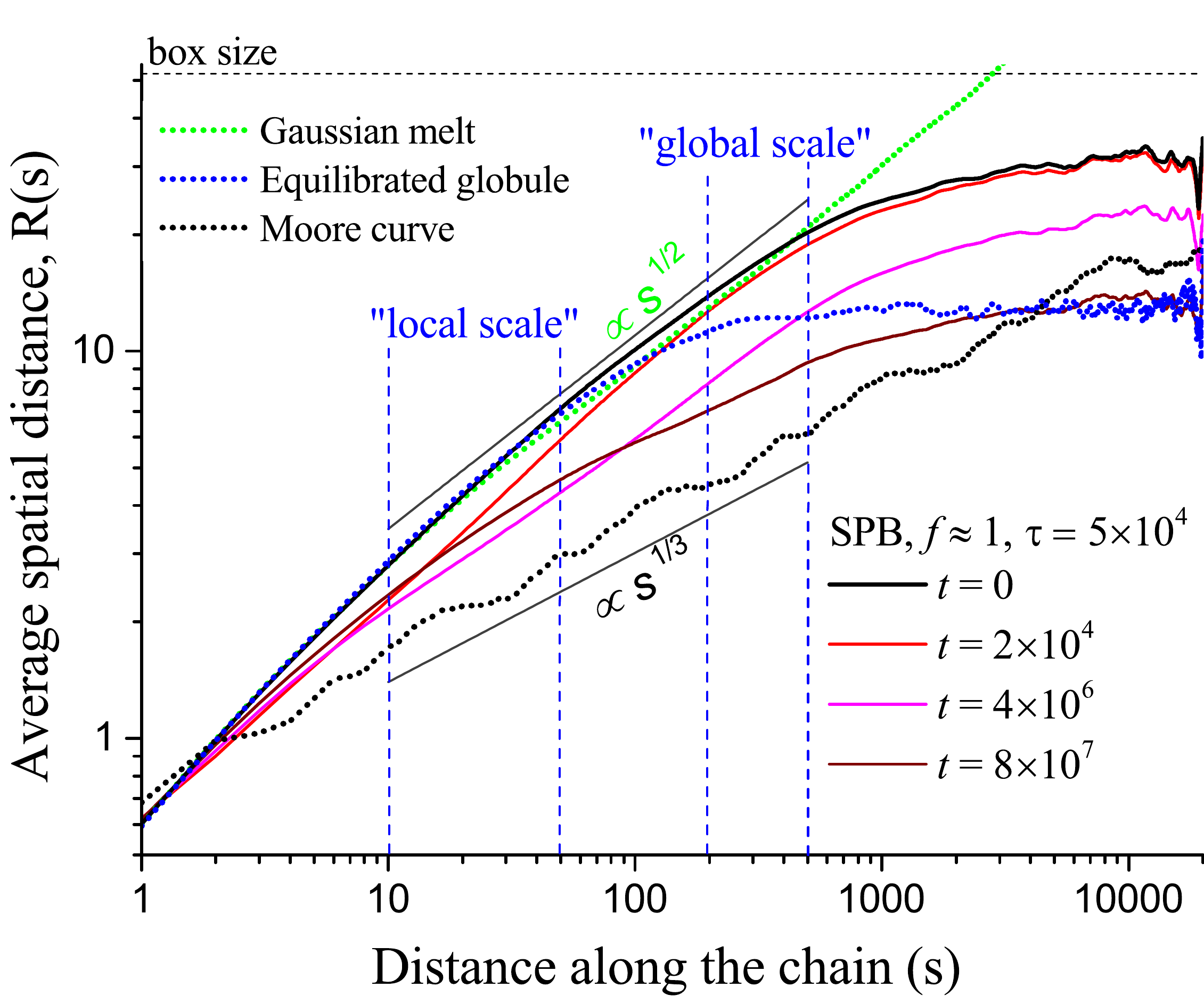}
	\caption{}
	\end{subfigure}
    \caption{$R(s)$ dependencies after $t$ DPD time steps: a) SPS in a very poor solvent ($\chi \approx 10$); b) SPB in athermal solvent with maximal fraction of bonds $f\approx 1$, bond lifetime $\tau = 5\times10^4$ DPD time steps. Green and black dotted lines represent dependencies for Gaussian (equilibrium) melt and Moore curve, respectively. $R(s)$ dependency for fully equilibrated globule is shown by the blue dotted line. Thin straight solid lines are for eye guide: black and green line represents $R(s)$ scaling for the fractal globule and Gaussian (equilibrium) melt, respectively. Chain length $N=2\times10^4$.}
    \label{rs_evolution_system_1_2}
\end{figure}

We studied a homopolymer chain (length $N=2\times10^4$ beads) in athermal solvent (which corresponds to zero Flory-Huggins parameter, $\chi$, or, equivalently, self-avoiding walk (SAW) in 3D space). We calculated $\chi$ value as $\chi = 0.306\times(a_{ij}-a_{ii})$ (Ref. \cite{groot1997dissipative}). Repulsion coefficients between solvent (type $1$) and polymer beads (type $2$) $a_{ij}$, $i,j = {1,2}$, were chosen to meet the criterion $\chi = 0$, and to make chain non-phantom \cite{nikunen2007reptational} ($a_{ij} = 150$, $i,j = {1,2}$, equilibrium bond length $r_0\approx 0.6$). To forbid self-intersections of the chain, we also set large bond stiffness parameter $K=150.0$, as described in Ref. \cite{nikunen2007reptational}.
Therefore, using such parameters of repulsion and bond stiffness, we are able to reproduce effects connected with topological constraints correctly.
Simulation box with impermeable boundaries was chosen to be a cube of $52\times52\times52$ DPD units, so the polymer concentration was $n \approx 0.06$ (DPD density $\rho=3$ particles per unit volume).
We decided to use simulation box with impermeable boundaries for several reasons. As we discuss in Results section, collapse induced by pairwise bonds formation is relatively slow. On the other side, if there are periodic boundary conditions (PBC), "ghost" beads repulse from "real" beads. This effect can be avoided only by choosing a very large simulation box. However, it is impossible to simulate a long chain in a very large simulation box due to explicit solvent presence (number of particles grows as the third power of the simulation box linear size). If collapse is rapid, we can neglect this nonphysical repulsion between “ghost” and “real” beads. However, if we study a slow collapse from a swollen state, there is sufficient time for the chain to affect its own conformation and dynamics. Therefore, we should take into account, how chain affects its own structure through PBC. This is a very challenging task, so we implemented a simpler solution: to use impermeable boundaries in the large enough simulation box. The polymer concentration was chosen to be low enough to let the chain swell in athermal solvent (Fig. \ref{general}, \ref{rs_evolution_system_1_2}, \ref{rs_system_1_2}). Structure of such a confined swollen state is well described by scaling theory of semi-dilute solutions, we discuss it later in Methods. Additionally, since our study is connected with chromatin organization description, it is natural to study processes in a confined space, because chromosomes are confined in a cell nucleus.
Information about other simulation parameters can be found in Supporting Information. We studied the process of collapse induced by pairwise bonds formation in the system described above. Pairwise reversible bonds, which are described by the potential for chain connectivity, can be formed with certain probability between any non-neighboring beads, which are spatially closer, than cutoff radius, $R_{c} = 1$. The detailed description of the algorithm of bond formation and breaking is given in Supporting Information. We note here that neglecting energy contribution in bond formation mechanism does not affect quantities measured in this work due to effective local thermostat in DPD (see Supporting Information for details). We define $N_{stp}$ as the number of DPD time steps between calculation of formation/breaking of bonds. It was equal to 200 DPD time steps in all our simulations. The average lifetime of a bond (hereafter: "bond lifetime") is calculated as the ratio $\tau = N_{stp}/\beta$, where $\beta$ is the probability of breaking a bond. To vary the bond lifetime, we varied $\beta$. Further we call this type of systems as System with Pairwise Bonds (SPB). Probabilities of bond formation and breaking set the time-averaged number of bonds formed inside the chain. We denote fraction of beads, which formed reversible bonds, as $f$. Hereafter we name this quantity as "fraction of bonds" for simplicity. 

We compared results for SPB with the process of collapse of the identical chain, placed in a poor solvent. Monomer-monomer and solvent-solvent repulsion parameters $a_{ii}$ $(i=1,2)$ were set as in SPB, but monomer-solvent repulsion was increased. We studied behavior of this system in a weak poor solvent ($\chi\approx1.5$) and in a very poor solvent ($\chi\approx10$), so the monomer-solvent repulsion parameters were increased up to $a_{12} = a_{21} = 154.5$ and $a_{12} = a_{21} = 183.0$, respectively. Other simulation parameters were set as in SPB. Further we call this system as System in a Poor Solvent (SPS).

Initial conformation from which the simulations started in SPB and SPS, was a swollen chain in athermal solvent inside the box with impermeable boundaries. To generate this structure, we placed chain with initially Gaussian statistics in athermal solvent conditions, and equilibrated the structure for $2.4\times10^7$ DPD time steps. We calculated Rouse time in good solvent $\tau_{R}$ to validate this simulation time. Mean-squared displacement of a bead $MSD(t)\propto t^{2/(2+d_{F})}$, where $d_{F}$ is the fractal dimension of the chain \cite{tamm2015anomalous} for Rouse subdiffusion regime. For a chain in athermal solvent $d_{F} \approx 5/3$, therefore, $MSD(t) \propto t^{6/11}$ and the Rouse time of the chain in good solvent is $\tau_{R} = \tau_{0}N^{2.2}$. From $MSD(t)$ measurements we obtained $\tau_{0} \approx 700$ - number of DPD time steps, sufficient for a bead diffusion on its size. Hence, the number of beads in a subchain, which has the Rouse time $\tau_{R} = 2.4\times10^7$, is $N_{s}\approx100$. This value has the same order of magnitude as the average number of beads inside a concentration blob in our system, $N_{conc.blob}\propto \Phi^{-5/4}\approx230$ ($\Phi = n\times a^3$ – volume fraction of polymer, $n$ is the concentration and $a\approx0.6$ is the size of a bead) \cite{de1979scaling}. Due to screening of volume interactions, spatially confined chain in athermal solvent is effectively Gaussian on a scale larger, than the concentration blob scale. Therefore, equilibration of a Gaussian chain for $2.4\times10^7$ DPD time steps is sufficient for formation of a fully equilibrated swollen chain in athermal solvent in a box with impermeable boundaries.

We developed the following method for studying kinetics of coil-globule transition. Let us consider dependencies of the average spatial distance between beads, separated by $s$ monomers ($R(s)$), after $t$ DPD time steps passed after the start of the simulation (we used 5 different initial conformations to obtain 5 $R(s)$ dependencies for each $t$, and built the averaged dependency, Fig. \ref{rs_evolution_system_1_2}). We see, that it is hardly possible to make conclusions and compare the processes of collapse, basing only on these plots. Therefore, we built linear approximations in log-log scale for each $R(s)$ dependency. This was made on the two scales: $s\in [10,50]$, further called "local" scale, and $s\in [200,500]$, further called "global" scale. We obtained information about how the scaling factor $\alpha_L$ in $R(s)\propto s^{\alpha_L}$ on the "local" scale, and the $\alpha_G$ in $R(s)\propto s^{\alpha_G}$ on the "global" scale change with time (Fig. \ref{slopes_system_1_2}, S8). We determined $\alpha_L$ and $\alpha_G$ for each of the 5 $R(s)$ dependencies for a given $t$. Then we plotted the average value and standard deviation of the scaling factors (Fig. \ref{slopes_system_1_2}, S8). Classical pearl-necklace mechanism describes chain collapse as sequential merging of collapsed "blobs". Therefore, if the pearl-necklace mechanism is realized, the chain collapses first on the "local" scale, and after this collapses on the "global" scale, and the time evolution of $\alpha_L$ and $\alpha_G$ has a characteristic shape (Fig. \ref{slopes_system_1_2}a, S8a). We compared, how $R(s)$ scaling depends on time on both scales during the two types of coil-globule transition: induced by pairwise bonds formation in athermal solvent and in a poor solvent. We chose the aforementioned values for "local" and "global" scale, based on our previous research on the coil-globule transition in a very poor solvent \cite{dynamicsofvolumecollapse} and the following reasons.

First, we explain our choice of the "local" scale. Initially, chain has equilibrated structure of a swollen chain in athermal solvent, confined in a box with impermeable boundaries. Basic scaling theory suggests, that a single homopolymer chain in a box with impermeable boundaries forms semi-dilute solution of subchains, lying between the boundaries, if polymer concentration is much less than unity \cite{de1979scaling}. The polymer concentration, chosen in the SPS and SPB, $n\approx0.06$, satisfies the aforementioned condition. Therefore, similarly to the semi-dilute solution, there are three characteristic scales in the initial swollen state: concentration blob ($R(s) \propto s^{3/5}$), effectively Gaussian subchain ($R(s) \propto s^{1/2}$, this scaling is clearly seen between "local" and "global" scales in the initial conformation (Fig. \ref{rs_evolution_system_1_2}, \ref{rs_system_1_2}), and the scale, where all linear correlations along the chain are vanished after reaching the impermeable boundaries of the simulation box ($R(s) \propto s^0$, Fig. \ref{rs_evolution_system_1_2}, \ref{rs_system_1_2}). Number of beads per concentration blob scales as $\Phi^{-5/4}$ ($\Phi^{-5/4}\approx 230$ in SPB and SPS). Therefore, $\alpha_L$ has a theoretically predicted value, close to $\alpha_L\approx3/5 = 0.6$ on the "local" scale prior to the collapse (Fig. \ref{slopes_system_1_2}, S8). Collapse of a homopolymer chain in a poor solvent starts with nucleation: formation of dense "blobs" along the chain \cite{lappala2013raindrop}. As dense "blobs" grow, the exponent $\alpha_L$ decreases. The value of the exponent reaches its minimum, when "blobs" have large enough size. Along with collapse, chain starts to locally expand in the already collapsed "blobs" (in other words, the chain starts to "mix"): polymer chain tends to form Gaussian conformations in polymer environment due to Flory theorem. The minimum value of $\alpha_L$ depends on how fast mixing occurs, and is lower in the chains with slow mixing with respect to collapse. After $\alpha_L$ reaches its minimum, it increases due to mixing in the already collapsed "blobs" up to the maximal value $\alpha_L=1/2$, corresponding to the equilibrium globule state.
Theoretical studies and computer simulation results show \cite{bunin2015coalescence,dynamicsofvolumecollapse,grosberg1988role,grosbergreview}, that chain segments, which form distinct blobs, remain segregated for a relatively long time and can not freely mix immediately after coagulation of blobs due to topological constraints. This happens because chain can not cross itself, and there are no open chain ends (except the two terminal ends), as in the polymer solution of many chains. Topological constraints, however, start to influence dynamics of the chain segments on the scale larger, than some characteristic length $\xi$, analogical to the tube diameter in linear chain solutions \cite{sakaue2011ring,rosa2014ring}. SPS in a very poor solvent is similar to the system studied in the work \cite{dynamicsofvolumecollapse}, therefore, characteristic maximum length of a polymer segment, which behavior is not affected by topological constraints, is $N_{\xi} \approx 50$ beads in SPS and SPB.
Therefore, it is convenient to study evolution of the exponent of $R(s)$ dependency on the scale $s<N_{\xi}$ to obtain information about structural evolution of unstable "blobs", i.e. about interplay between chain mixing and collapse on the local scale, without influence of topological constraints between non-phantom segments of the same chain. We note, however, that due to effective excluded volume of chain beads, the beginning of $R(s)$ dependency is steeper, than the remaining dependency. This behavior is universal for all studied chains (Fig. \ref{rs_evolution_system_1_2}, \ref{rs_system_1_2}) up to the scale $s\approx10$. Therefore, it is natural to describe evolution of the local structure of the chain by the exponent $\alpha_{L}$ in the dependency $R(s)\propto s^{\alpha_{L}}$, determined on the scale $s\in[10,50]$.

Second, we determined the "global" scaling factor $\alpha_G$ to characterize kinetics of collapse on a scale $s>N_{\xi}$, where topological constraints start to affect mixing process. Plateau in $R(s)$ dependency in equilibrium globule roughly starts at $s\approx 200$ scale (Fig. \ref{rs_evolution_system_1_2}, \ref{rs_system_1_2}). Therefore, it is convenient to determine $\alpha_G$ on the scale $s > 200$, because $\alpha_G\approx 0$ in the equilibrium globule state. Therefore, if $\alpha_G$ is sufficiently larger than zero, and this behavior is stable in time, we observe formation of a metastable non-equilibrium structure. We can see, that impermeable boundaries start to affect the $R(s)$ dependencies in the initial swollen state on the scale $s\approx500$ (Fig. \ref{rs_evolution_system_1_2}, \ref{rs_system_1_2}). Hence, $\alpha_G$ value before collapse is slightly smaller than the expected value $1/2$, which should be observed in the semi-dilute solution (Fig. \ref{slopes_system_1_2}, S8).

We also studied effect of pairwise reversible bonds presence on subdiffusion inside crumpled globule (CG) and equilibrium globule (EG). To obtain the CG and EG, we first generated fractal and equilibrium globule structure as a single Moore curve and a random walk, respectively. Number of beads in CG and EG is $N = 32768$, each globule was placed in a simulation box with impermeable boundaries sized $21.4942\times21.4942\times21.4942$ DPD units (it means, polymer concentration is $n\approx 1$, DPD density $\rho=3$ particles per unit volume). Then we annealed the generated structure for $t_{0}=4\times10^{7}$ DPD time steps. The structure relaxes after annealing, but $t_0$ is small enough, and the crumpled globule can not form knots \cite{tamm2015anomalous}. After the aforementioned procedure we measured $MSD(t)$ - the time dependency of mean-squared displacement of a bead. Exact procedure of $MSD(t)$ calculation is given in the Supporting Information. 

\section{Results}

\subsection{Kinetics of coil-globule transition}

\begin{figure}[h!]
    \centering
    \begin{subfigure}{0.49\textwidth}
		\includegraphics[width=\linewidth,height=\textheight,keepaspectratio]{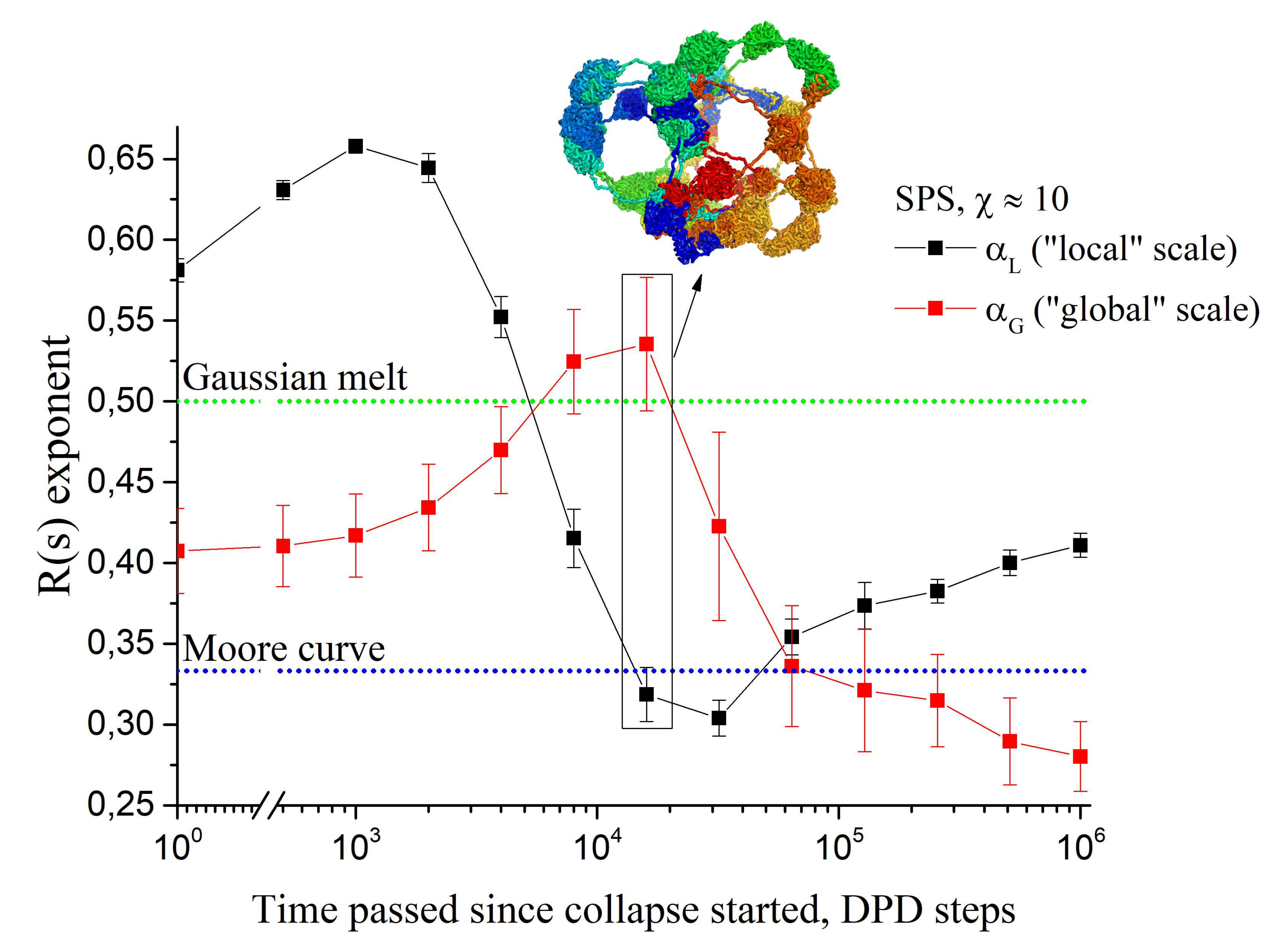}
		\caption{}
		\end{subfigure}
		\begin{subfigure}{0.49\textwidth}
		\includegraphics[width=\linewidth,height=\textheight,keepaspectratio]{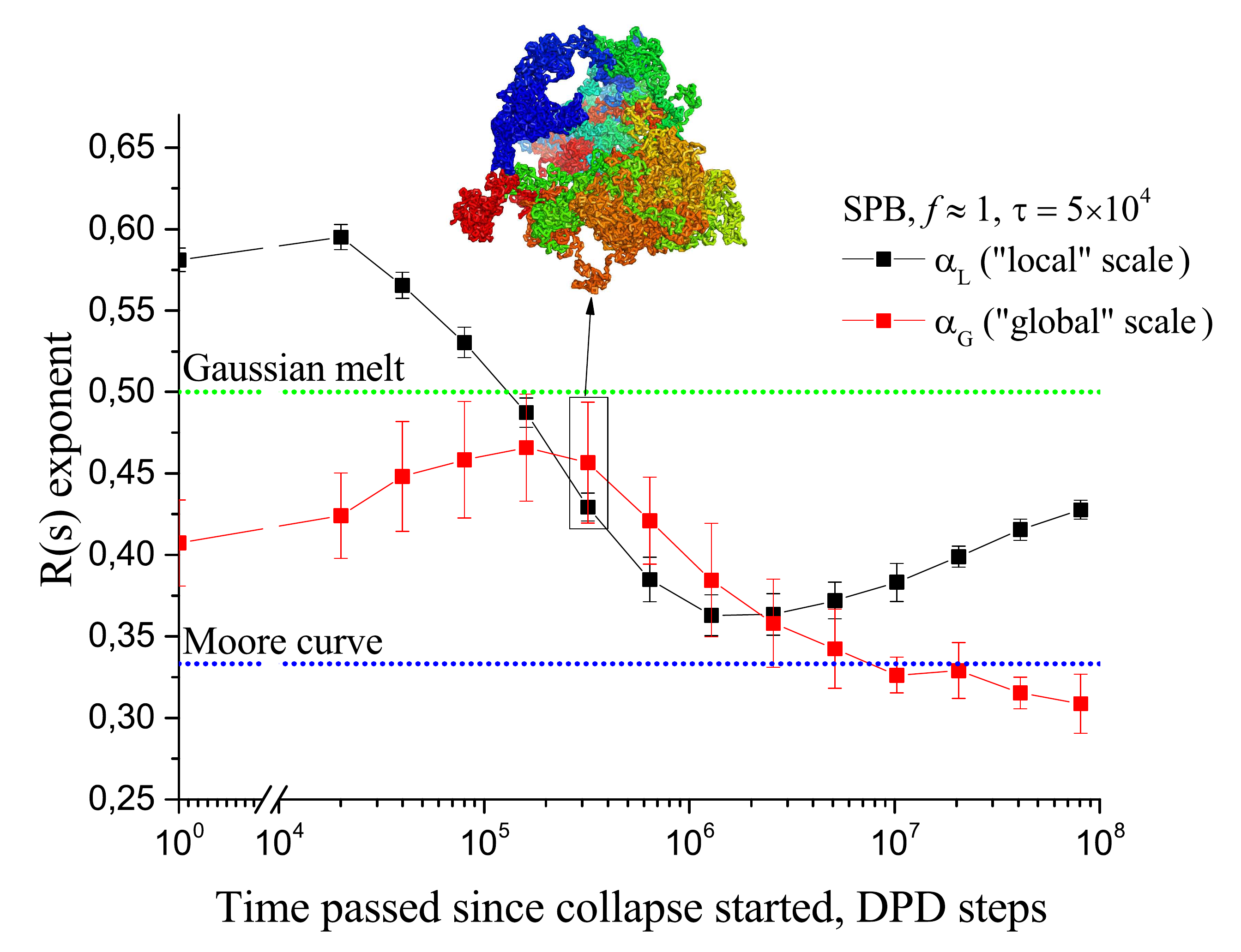}
		\caption{}
		\end{subfigure}
    \caption{"Local" ($\alpha_L$) and "global" ($\alpha_G$) $R(s)$ scaling exponents for System in a Poor Solvent (SPS, $\chi\approx10$, Figure a) and System with Pairwise Bonds (SPB, fraction of bonds $f\approx 1$, bond lifetime $\tau = 5\times10^4$ DPD steps, Figure b), depending on the time passed since simulation started.}
    \label{slopes_system_1_2}
\end{figure}

First we studied kinetics of collapse in a poor solvent (SPS) as a reference case. We present results for a chain behavior in a very poor solvent in the Manuscript ($\chi \approx 10$, Fig. \ref{rs_evolution_system_1_2}a, \ref{slopes_system_1_2}a) and in a weak poor solvent in the Supporting Information ($\chi \approx 1.5$, Fig. S8a). As we expected, collapse of the chain in a poor solvent follows the pearl-necklace mechanism. The chain tends to form "blobs" on the "local" scale, as described in Ref. \cite{dynamicsofvolumecollapse}. Hence, after small increase $\alpha_L$ decreases and reaches its minimal value (i.e. the chain becomes fully collapsed on the scale $s<N_\xi$) on the time scale $t\propto 10^4$ DPD time steps in a very poor solvent and on the time scale $t\propto 10^5$ DPD time steps in a weak poor solvent. As chain collapses on the local scale, it starts to resemble a swollen chain on the large scale, as predicted by the pearl-necklace model of collapse in a poor solvent. Therefore, we observe increase of the exponent $\alpha_G$ on the "global" scale. As we see, chain remains swollen up to $t\propto 10^4$ DPD time steps in a very poor solvent and up to $t\propto 10^5$ DPD time steps in a weak poor solvent on the "global" scale (Fig. \ref{slopes_system_1_2}a, S8a). Then the chain collapses on the "global" scale: $\alpha_G$ decreases. Therefore, classical pearl-necklace mechanism of collapse is observed: chain collapses first on a small scale, and only then it collapses on a larger scale. Typical conformations of the chain prior to the collapse on the "global" scale are included in the Fig. \ref{slopes_system_1_2}a, S8a, demonstrating classical pearl-necklace structures. It is worth to mention, that sufficiently large stable $\alpha_G$ value is observed after collapse (Fig. \ref{slopes_system_1_2}a, S8a). Therefore, metastable crumpled globule state forms, as described in \cite{dynamicsofvolumecollapse}.

\begin{figure}[h!]
	\includegraphics[width=0.49\linewidth,height=\textheight,keepaspectratio]{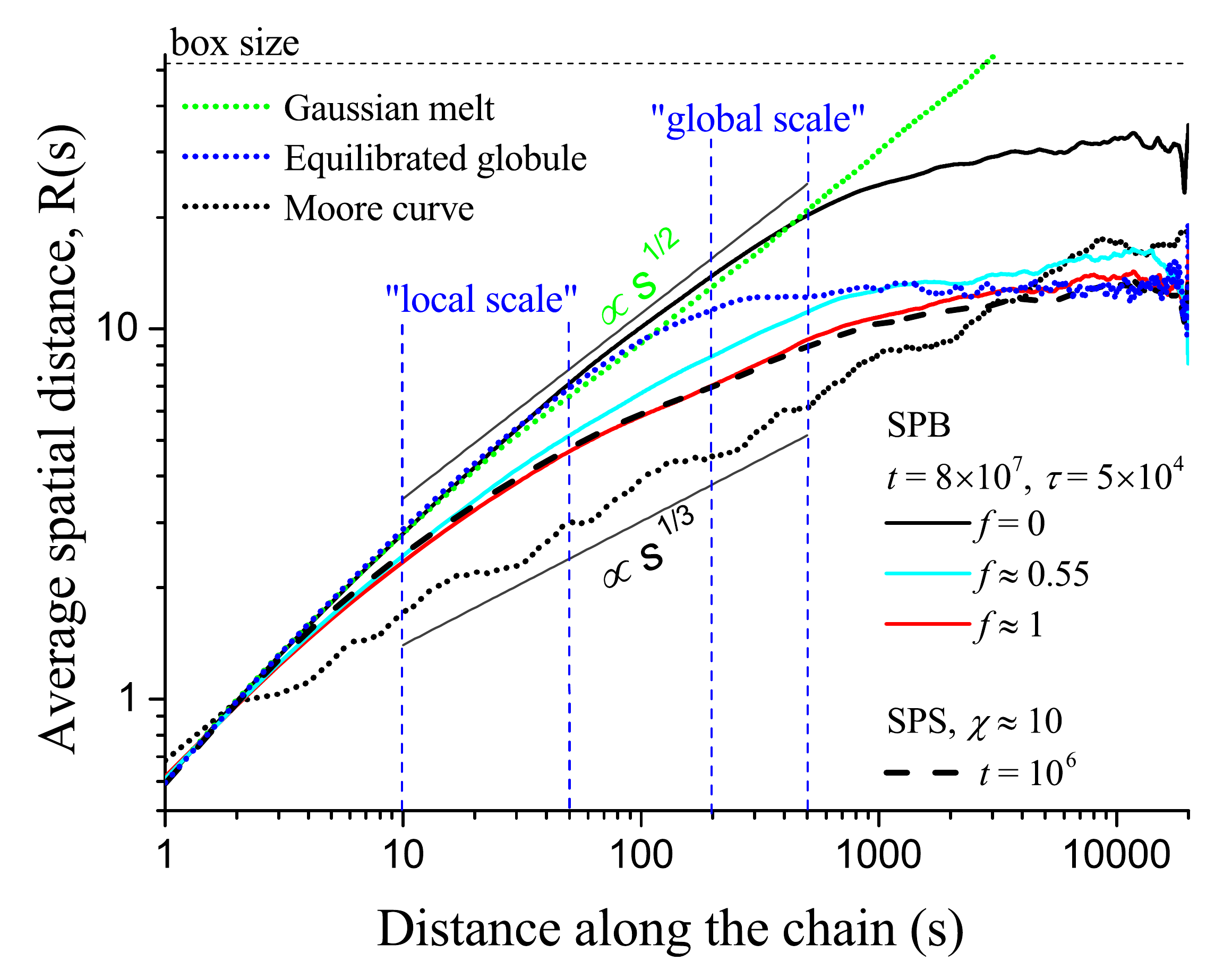}
    \caption{$R(s)$ dependencies for the steady-state regimes for SPB (with different fraction of bonds $f$) and SPS. In addition, data for fully equilibrated globule (blue dotted line) and Moore curve (black dotted line) are presented. Thin straight solid lines are for the eye guide: black and green line represents $R(s)$ scaling for the fractal globule and Gaussian (equilibrium) melt, respectively.}
    \label{rs_system_1_2}
\end{figure}

Then we studied collapse, induced by pairwise bonds formation (SPB). As we see from Fig. S9, radius of gyration of the chain in the SPB decreases with increase of the average fraction of bonds, coil-globule transition occurs roughly at $f\approx0.3$. First we studied kinetics of coil-globule transition of the chain with maximal fraction of bonds $f\approx 1$ (probability of bond formation $=1$), bond lifetime was set to $\tau = 5\times10^4$ DPD time steps (probability of bond breaking $\beta=0.004$), which corresponds to the maximum influence of pairwise bonds. Fraction of bonds is approximately equal to its equilibrium value roughly after $2\times10^3$ DPD time steps passed after bond formation started (Fig. S10a). However, the chain remains almost swollen on both "local" and "global" scale: chain can not collapse during relatively fast bonds formation after $2\times10^3$ DPD time steps. Hence, collapse in such a system is induced by rearrangement of reversible bonds, followed by the gradual collapse of the chain. Figures \ref{rs_evolution_system_1_2}b, \ref{slopes_system_1_2}b demonstrate, that first slight collapse on a "local" scale and increase of $\alpha_G$ up to $t\approx 4\times10^4$ DPD time steps occurs similarly to collapse in a poor solvent. However, after that $\alpha_G$ reaches plateau, and after $t\approx3\times10^5$ DPD time steps starts to decrease simultaneously with $\alpha_L$. This means, that prior to the collapse on the "global" scale chain is not fully collapsed on the "local" scale and does not resemble the pearl-necklace structure (Fig. \ref{slopes_system_1_2}b). Collapse on the scale $s<N_\xi$ occurs simultaneously with collapse on the much larger scale. This situation is not observed for collapse in a poor solvent and is not typical for pearl-necklace mechanism of collapse. After $t\propto 10^6$ DPD time steps chain is fully collapsed on the "local" scale, and is sufficiently collapsed on the "global" scale, in contrast to the collapse in a poor solvent. After that chain starts to mix on the "local" scale, therefore, $\alpha_L$ starts to increase slowly. At the end of simulation time SPB formed a long-living crumpled globule state very similar to SPS case with the value $\alpha_L\approx 0.4$ and $\alpha_G\approx0.3$, compare the red and the dashed line in Fig. \ref{rs_system_1_2}.

In addition we studied kinetics of collapse in SPB, induced by reversible bonds formation with larger lifetime $\tau = 5\times10^5$ DPD time steps ($\beta=0.0004$), $f\approx1$. Bond rearrangement is slowed down stronger, as bond lifetime is increased,. Therefore, full collapse on the "local" scale occurs on the time scale up to $t \propto 8\times10^7$ steps (Fig. S8b). Collapse on the "global" scale slows down as well, but again proceeds simultaneously with collapse on the "local" scale, as previously described. Moreover, this effect is even stronger, than in the SPB with bond lifetime $\tau = 5\times10^4$ DPD time steps: $R(s)$ exponent on the "global" scale never exceeds the $\alpha_L$ value and decreases simultaneously with $\alpha_L$ much longer.

Finally, if probability of bond formation is set to $0.001$ and bond lifetime is $\tau = 5\times10^4$ DPD time steps, the equilibrium fraction of bonds $f\approx 0.55$ is reached after $t\approx2\times10^6$ DPD time steps (Fig. S10b). This is a very "soft" regime of collapse. Collapse is still induced by reversible bonds formation, but the mechanism of collapse is not based only on rearrangement of reversible bonds, but also on gradual increase of the fraction of reversible bonds in time. The effect of simultaneous collapse on the "local" and "global" scale is also observed (Fig. S8c). Snapshot of a typical chain conformation prior to collapse on the "global" scale illustrates the absence of pearl-necklace conformation. On the large time scale chain reaches the metastable globular state, similar to crumpled globule, as in the previously discussed cases.

\begin{figure}[h!]
	\includegraphics[width=0.49\linewidth,height=\textheight,keepaspectratio]{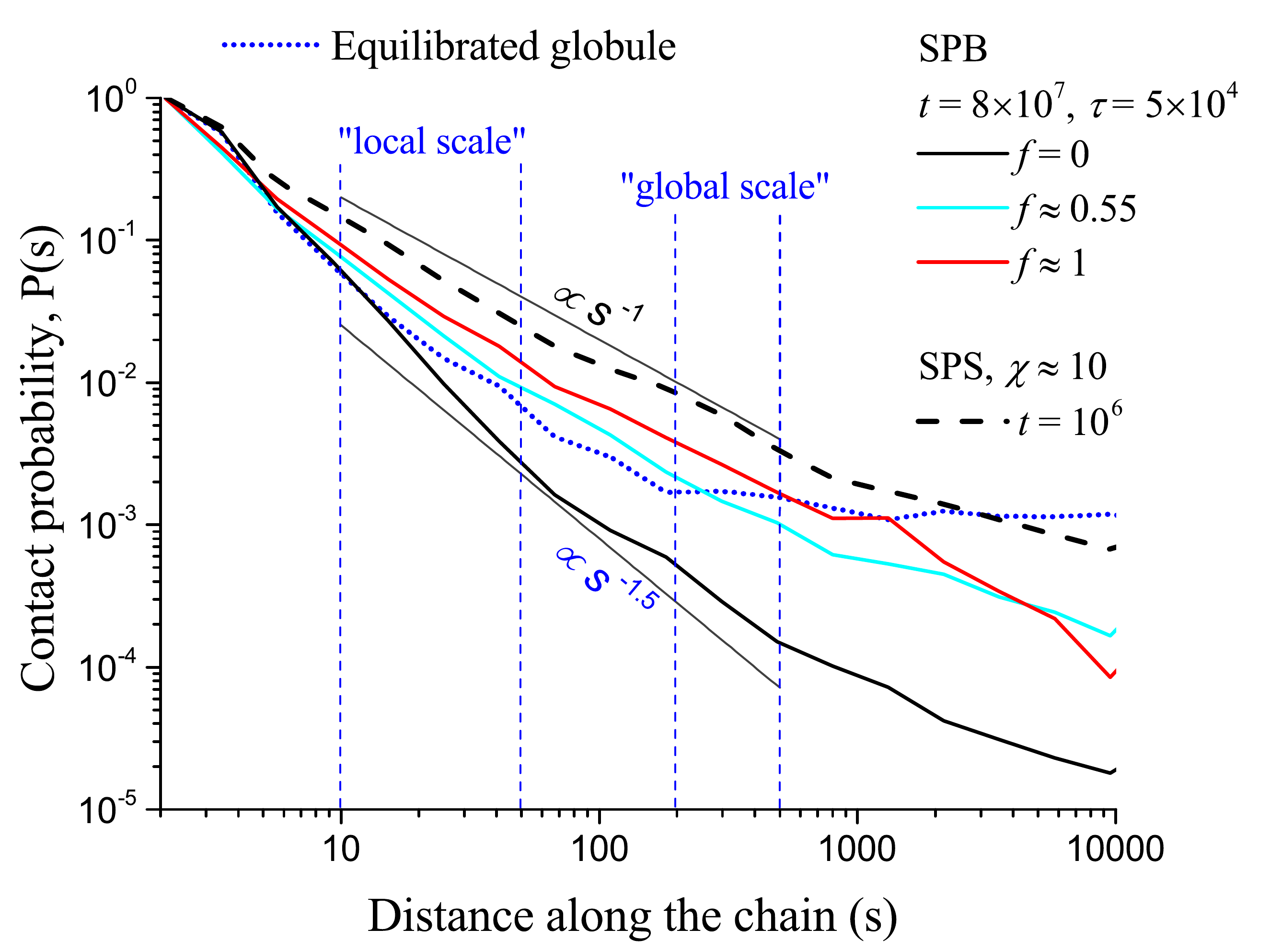}
	\caption{$P(s)$ dependencies for the steady-state regimes for SPB (with different fraction of bonds $f$) and SPS. SPB with $f\approx 1$ (red solid line) is approximated by $P(s)\propto s^{-1.12\pm 0.02}$ on the "local" scale and $P(s)\propto s^{-1.10\pm 0.06}$ on the "global" scale. Thin straight solid lines are for eye guide: black and blue lines represent $P(s)$ scaling for the fractal globule and Gaussian (equilibrium) melt, respectively.}
	\label{ps_system_1_2}
\end{figure}

We also analyzed probabilities of beads contact, depending on the separation along the chain, $P(s)$ (full information about $P(s)$ calculation is given in the Supporting Information). This is the simplest representation of data on chromosome conformations in typical Hi-C biological experiments. We studied the $P(s)$ dependency in the steady-state regime, after $t=8\times10^7$ DPD time steps. Similar to the $R(s)$ dependencies, we used 5 initial conformations to obtain 5 $P(s)$ dependencies for each state, and plotted the averaged data, Fig. \ref{ps_system_1_2}. We see, that the $P(s)$ scaling in the steady-state is very similar to the $P(s)\propto s^{-1}$ dependency for the fractal globule \cite{Hi-C}. The structure is slightly mixed on the "local" and on the "global" scale, and the $P(s)$ slope in the log-log scale appears to be somewhat smaller, than $-1$, but much larger than $-1.5$ on the "local" scale and much smaller than $0$ on the "global" scale, as in equilibrium globule (Fig. \ref{ps_system_1_2}). The structure with less fraction of bonds is mixed stronger, but is still far from reaching the $P(s)$ dependency for equilibrium globule. Therefore, $P(s)$ dependency confirms, that the metastable state, forming after collapse in a very poor solvent and in the SPB (bond lifetime is $\tau = 5\times10^4$ DPD time steps), resembles crumpled globule state.

\subsection{Dynamical properties}
As a result of the previous section, we observed, that a homopolymer chain forms a non-equilibrium metastable state, similar to crumpled globule during coil-globule transition. In this section we studied, how presence of pairwise reversible bonds affects subdiffusion in two compact states: Crumpled Globule (CG) and Equilibrium Globule (EG).

\begin{figure}[h!]
    \centering
    \begin{subfigure}{0.49\textwidth}
		\includegraphics[width=\linewidth,height=\textheight,keepaspectratio]{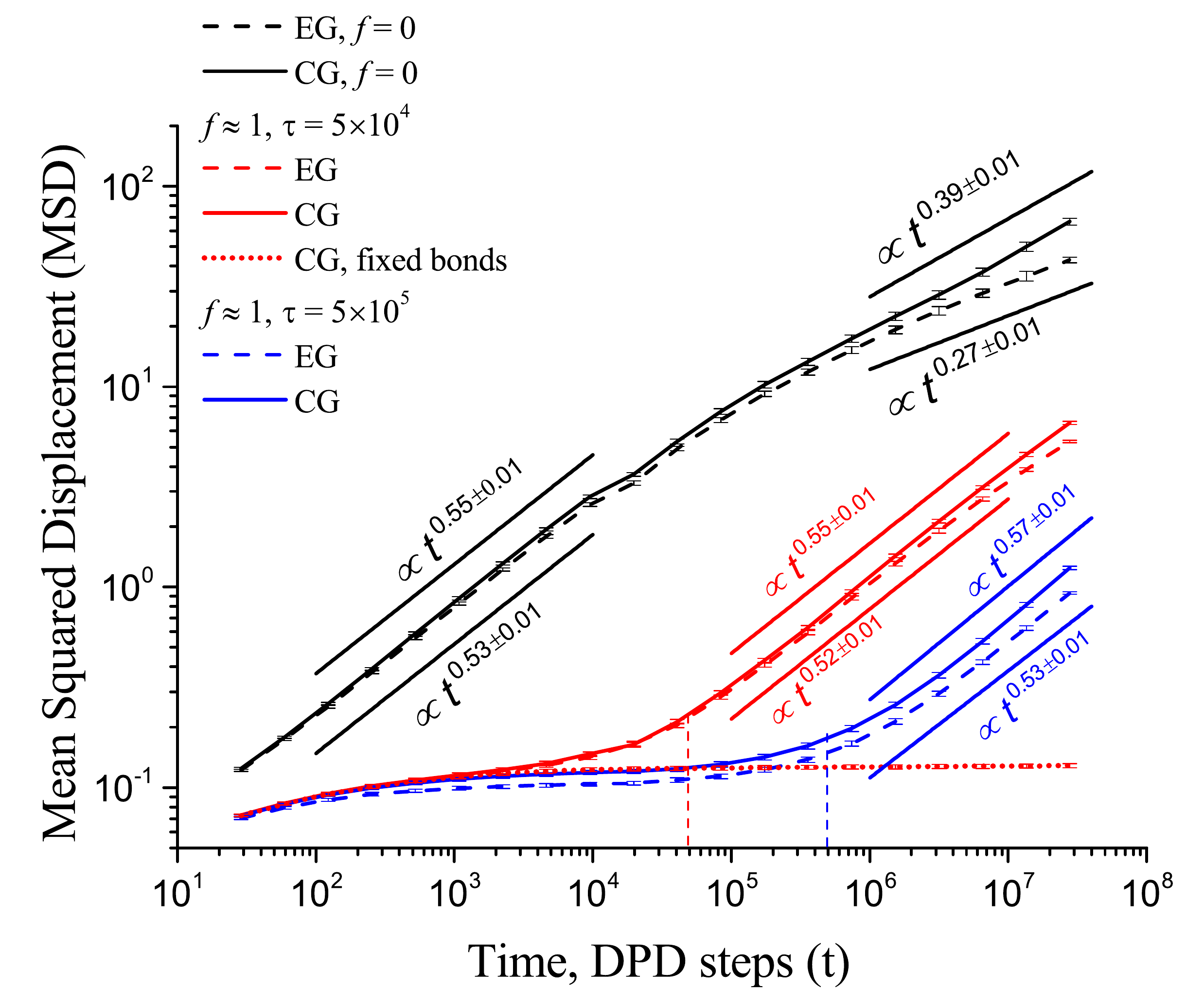}
		\caption{}
		\end{subfigure}
		\begin{subfigure}{0.49\textwidth}
		\includegraphics[width=\linewidth,height=\textheight,keepaspectratio]{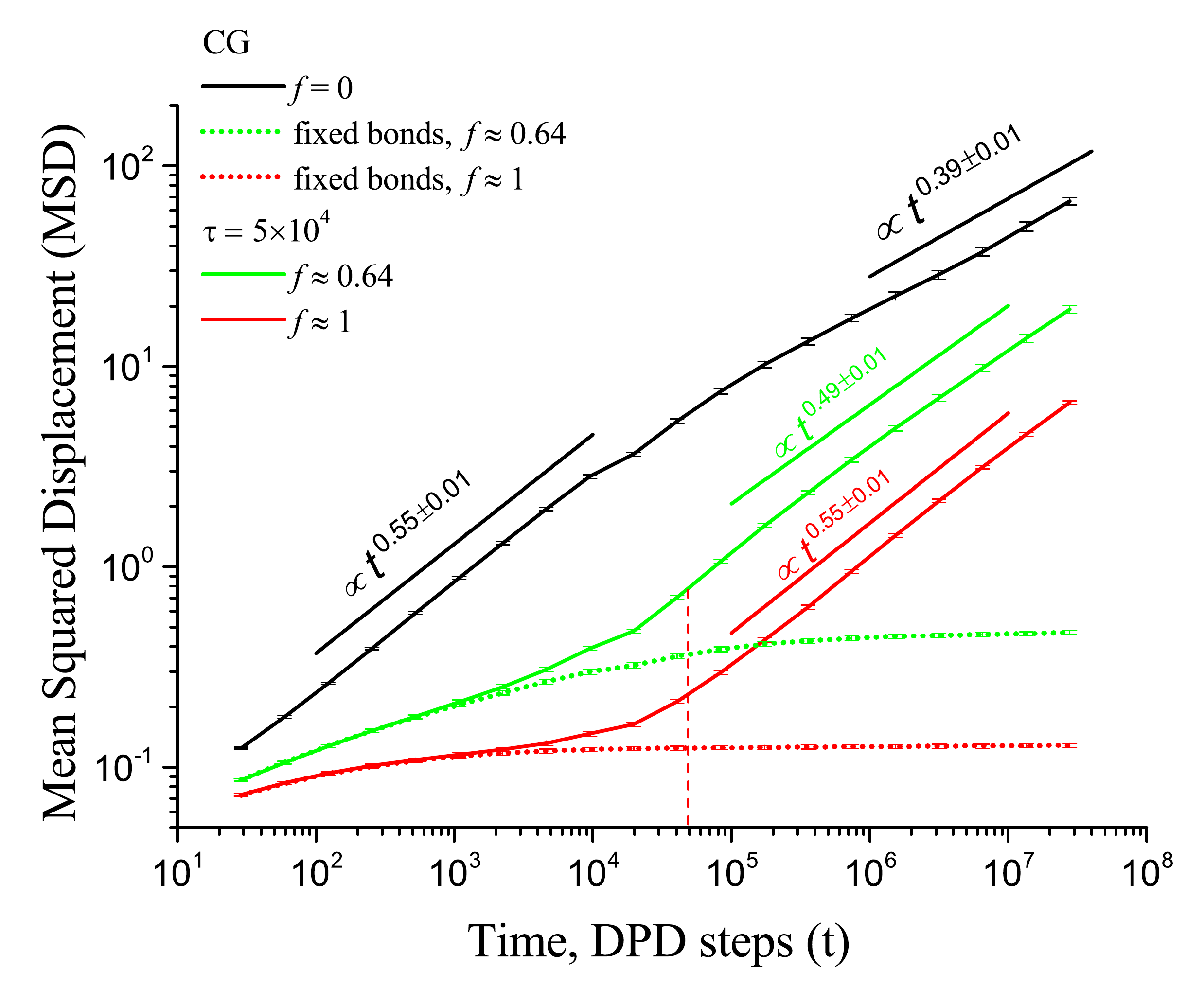}
		\caption{}
		\end{subfigure}
    \caption{$MSD(t)$ in Crumpled Globule (CG), solid lines, and for Equilibrium Globule (EG), dashed lines. Figures (a) and (b) demonstrate the influence of bond lifetime and fraction of bonds, respectively, on subdiffusion. Red and blue vertical dashed lines represent bond lifetime $\tau = 5\times10^4$ and $\tau = 5\times10^5$ DPD time steps, respectively. Green and red dotted lines represent $MSD(t)$ for CG with bond formation/breaking switched off after $t=2\times10^4$ DPD time steps; fraction of bonds is $f\approx0.64$ and $f\approx1$, respectively. Thin solid lines represent exact linear fitting of $MSD(t)$ dependency on the given time scale.}
    \label{msd_melts}
\end{figure}

First, we chose CG and EG without bonds as reference states, and measured $MSD(t)$ dependencies in both globules (black lines, Fig. \ref{msd_melts}). We see, that EG behaves dynamically as predicted by the tube model \cite{de1979scaling} due to the presence of knots in equilibrium globule \cite{virnau2005knots,imakaev2015effects}. CG behaves dynamically according to the modified Rouse model in fractal globules \cite{tamm2015anomalous}. Then we studied subdiffusion in both states with maximal fraction of bonds ($f\approx1$) of two lifetimes $\tau = 5\times10^4$ and $\tau = 5\times10^5$ DPD time steps (Fig. \ref{msd_melts}a). We see, that subdiffusion in both globules is "frozen" on the time scale shorter than bond lifetime. In other words, $MSD(t)$ dependencies for CG with reversible bonds and with "fixed" bonds coincide (to "fix" bonds, we switched off formation and breaking of bonds in CG after $2\times10^4$ DPD time steps and measured dynamics in this "frozen" system, see dotted lines in Figures \ref{msd_melts}a, \ref{msd_melts}b). However, on the time scale larger, than bond lifetime, dynamical behavior is well described by Rouse model in both EG and CG. In other words, dynamic behavior of CG and EG with $f\approx1$ is similar to subdiffusion in the CG and EG without bonds, respectively, on the time scale larger than bond lifetime. We could not reach $MSD(t) \propto t^{0.38}$ and $MSD(t) \propto t^{0.25}$ scaling regimes for CG and EG with $f\approx1$, respectively, due to computational restrictions.

Second, we analyzed, how dynamical behavior in the crumpled globule state changes with $f$. We measured $MSD(t)$ for CG with $f\approx0.64$, bond lifetime equal to $\tau = 5\times10^4$ DPD time steps (Fig. \ref{msd_melts}b). We observed that on the time scale shorter than bond lifetime subdiffusion coincides with $MSD(t)$ for CG with "fixed" bonds and varies with $f$. Dynamics becomes faster as the fraction of bonds decreases. We suppose, that dynamics in such systems depends on the average density of reversible bonds on the time scale shorter than the bond lifetime. Density of bonds is directly related to fraction of bonds, and controls the bead motion damping rate. Decrease of $f$ leads to decrease in damping rate, and, therefore, to increase of average bead mobility. Dynamical behavior of CG on the time scale larger, than bond lifetime, does not depend on $f$ and it is well described by Rouse model.

Our results on subdiffusion in compact states with presence of pairwise reversible bonds may be used to assess the stability of the crumpled globule formed during coil-globule transition. Increase of bond lifetime and fraction of bonds leads to hampered subdiffusion on the larger time scale. Therefore, it results in slower mixing of the chain. Hence, we suggest, that existence time of the aforementioned metastable state can be controlled by the fraction of bonds and bond lifetime.

\section{Discussion}
In this work we have studied structure of transient states in coil-globule transition induced by pairwise reversible bonds formation. We found out that this type of collapse was not kinetically similar to collapse in a poor solvent. We stress that this difference is observed due to finite bond lifetime in the SPB. Finite bond lifetime leads to slow rearrangement of pairwise bonds, so the chain relaxes between two consecutive rearrangements. The absence of the chain relaxation during collapse is crucial for pearl-necklace structure formation. Increase of the bond lifetime prolongs the relaxation of the chain between bond rearrangements. Therefore, collapse on the "global" scale occurs simultaneously with collapse on the "local" scale. Decrease of equilibrium fraction of bonds in the chain does not affect the qualitative picture. The transient structures formed prior to collapse on the "global" scale are still not of the pearl-necklace type. Therefore, difference in coil-globule transition kinetics is controlled by the type of interaction between beads.

In addition we found out, that steady-state collapsed states were very similar in both types of coil-globule transitions under study. The steady-state resembles crumpled globule state, which is one of the well-known models of interphase chromatin structure in different biological species. Stability of this state can be controlled by fraction of bonds and bond lifetime, as it was initially suggested in the work \cite{schram2013stability}.

As we mentioned in Introduction, different models are aimed to describe different scales and properties of chromatin. There is a well-defined border between inter-TAD and intra-TAD chromatin organization: $P(s)$ scaling inside TADs is significantly different from $P(s)$ on a larger scale \cite{sanborn2015chromatin,drosophilaconf}. We propose here a novel model for large-scale chromatin organization. We choose TAD as a basic interacting unit. We describe chromatin as a coarse-grained flexible homopolymer chain in athermal solvent. One bead in our model chain roughly represents a single TAD. The reversible bonds are treated as relatively long-living inter-TAD contacts (we consider them as pairwise interactions and neglect triple and more complex inter-TAD contacts). There are several biological mechanisms underlying this interactions, for example non-acetylated histone H3 tail forming a complex with acidic region on H2A-H2B histone dimer \cite{histoneinteractions}, or interactions through insulator proteins, such as cohesin \cite{nasmyth2001disseminating,loopextrusion}. We emphasize the reversible pairwise nature of the aforementioned interactions, so we introduce this type of bonds in our model. In general, pairwise reversible interactions are abundant in all living systems: sulfide bridges and hydrogen bonds play important role in protein folding and DNA double-stranded structure formation. There are a lot of mechanisms of pairwise reversible interactions in chromatin, depending on the type of chromatin \cite{jost2014modeling}.

$P(s) \propto s^{-1}$ behavior, which is characteristic of our steady-state crumpled globule state (Fig. \ref{ps_system_1_2}) was observed in mammals and Drosophila (Fig. S5) \cite{barbieri2012complexity,grosbergreview}. Also in the work \cite{tadspositions} the transition from $R(s) \propto s^{0.33}$ to $R(s) \propto s^{0.2}$ was observed on the $\approx 7 Mbp$ scale, similarly to the $R(s)$ behavior in crumpled globule state (Fig. \ref{rs_system_1_2}). Similar $R(s)$ scaling was observed in human male fibroblast cells \cite{munkel1999compartmentalization}. By changing the number and lifetime of pairwise reversible inter-TAD contacts, one can reproduce different dynamical properties of chromatin on a small time scale. On a large time scale, $MSD(t) \propto t^{0.4}$ dependency is observed in different biological species \cite{telomeres1,kepten2013improved,weber2012nonthermal} (Fig. \ref{msd_melts}). Our crumpled globule state exists in a rather wide range of the fraction of bonds ($f\in[0.55;1]$), therefore, the model is resistant to fluctuations of parameters. Moreover, $MSD(t)$ measurements suggest (Fig. \ref{msd_melts}), that stability of the crumpled globule state, formed after collapse in SPB, can be controlled by two parameters: fraction of bonds and bond lifetime. Therefore, the suggested model is more or less universal and can explain several slightly different genome organizations in various biological species.

We could not reproduce ensemble-averaged experimental Hi-C maps with our homopolymer chain based model, because we neglected the difference between interactions of different types of chromatin \cite{jost2014modeling}. Moreover, rather simple confinement conditions in our model can not reproduce effects, connected with sophisticated lamina structure, for example, tethering of a chromosome at specific cites \cite{gasser2002visualizing}. Development of the heteropolymer model with pairwise reversible bonds of different types, which can lead to the microphase separation, as well as to formation of the active and inactive chromatin compartments \cite{Elgin1996,tadspositions,drosophilaconf}, and to effects connected with lamina structure, is on the way \cite{chertovichcopolymermodeldrosophila,ulianov2019nuclear} and is the matter of our next study.

\section{Acknowledgments}
We thank M.V. Tamm for fruitful discussions and comments, A.A. Gavrilov for providing DPD computer code and A.A. Galytsyna for the experimental data and useful remarks. The research of Pavel Kos and Artem Petrov is supported partly by Skoltech Systems biology fellowship and grant from the Foundation for the advancement of theoretical physics ``Basis''. The reported study was funded by RFBR according to the research project \# 18-29-13041.

\section{References}
\bibliography{ref}

\renewcommand{\thefigure}{S\arabic{figure}} 
\renewcommand{\theequation}{Eq. S\arabic{equation}}

\section{Supporting Information}

\subsection{Calculation of $P(s)$ and $R(s)$}
To calculate $P(s)$ we counted $N_{i}$ - the number of beads, which have less distance to the $i$-th bead, than cutoff radius = 1.5, and then averaged $N_{i}$ over all measurements in a chain:

\begin{equation}
\label{eq:ps}
    P(s) = \frac{1}{N-s} \sum_{i=1}^{N-s}N_{i}(t_{0})
\end{equation}

In $R(s)$ we calculated average spatial distance between beads, separated by s beads along the chain.
\begin{equation}
\label{eq:rs}
\begin{split}
    & R(s) = \frac{1}{N-s} \sum_{i=1}^{N-s}((x_{i}(t_{0})- \\
    & -x_{i+s}(t_{0}))^{2}+(y_{i}(t_{0})-y_{i+s}(t_{0}))^{2}+ \\ 
    & +(z_{i}(t_{0})-z_{i+s}(t_{0}))^{2})^{1/2}
\end{split}
\end{equation}

\subsection{Procedure of $MSD(t)$ calculation}
We calculated $MSD(t)$ according to the following procedure. Using the annealed structure as the initial structure, we continued the simulations, and structures were generated every $\Delta t$ DPD time steps. However, we chose two different time steps: $\Delta t_1 = 10$ DPD time steps and $\Delta t_2 = 2\times10^4$ DPD time steps. Simulations, where structures were generated every $\Delta t_1$ DPD time steps, were performed during $T_1 = 2\times10^4$ DPD time steps, and, where structures were generated every $\Delta t_2$ DPD time steps, were performed during $T_2 = 4\times10^7$ DPD time steps. This was done for computational efficiency: we needed to measure $MSD(t)$ in a very broad time range $t\in[10,4\times10^7]$, and if we give out structures with one time step $\Delta t=10$, it would not be possible to measure $MSD(t)$ in a reasonable time due to extraordinarily large number of structures to analyze. Therefore, we "split" the time range into two parts, and gave out structures with two different time steps. We ensured the correctness of such approach by calculating $MSD(t=2\times10^4)$ and verifying, that both dependencies give the same value and the standard deviation of mean squared displacement.

The program analyzed n-th pair of structures, separated by $t=n\Delta t_k$, $(k=1,2)$ DPD time steps from each other, counting $MSD_n(t)$: the average squared displacement of beads in the given pair of structures. We also distracted displacement of the center of mass of the chain (COM) from displacements of beads. Then program analyzed next pair of structures, separated by $t$ DPD time steps from each other, and calculated $MSD_{n+1}(t)$ etc. $MSD(t)$ was calculated as the average over all $MSD_n(t)$ values. Standard deviation of $MSD(t)$ was calculated as the standard deviation in the set of $MSD_n(t)$ values. Exact equation for $MSD(t)$ calculation is given in Equation \ref{eq:msd}.

\begin{equation}
\label{eq:msd}
    \begin{split}
        & MSD(t, t<T_k-t_0) = \frac{1}{N} \frac{\Delta t_k}{T_k-(t+t_0)}\sum_{n=0}^{(T_k-(t+t_0))/\Delta t_k}\sum_{i=1}^{N}((x_{i}(t+t_{0}+n\Delta t_k)- \\ 
        & -x_{COM}(t+t_{0}+n\Delta t_k))-(x_{i}(t_{0}+n\Delta t_k)-x_{COM}(t_{0}+n\Delta t_k))^{2}+ \\ 
        & +((y_{i}(t+t_{0}+n\Delta t_k)-y_{COM}(t+t_{0}+n\Delta t_k))-(y_{i}(t_{0}+n\Delta t_k)- \\
        & -y_{COM}(t_{0}+n\Delta t_k))^{2}+ \\ 
        & +((z_{i}(t+t_{0}+n\Delta t_k)-z_{COM}(t+t_{0}+n\Delta t_k))-(z_{i}(t_{0}+n\Delta t_k)- \\
        & -z_{COM}(t_{0}+n\Delta t_k))^{2}, k=1,2
    \end{split}
\end{equation}

\subsection{Bond formation/breaking algorithm}
In this section we describe in detail, how formation and breaking of pairwise reversible bonds is realized in our simulations. Every $N_{stp} = 200$ DPD time steps program chooses an i-th bead, and forms a list of beads, which are spatially closer to the chosen i-th bead, than cutoff radius $R_c = 1.0$. Then this list is sorted by spatial distance to the i-th bead (beads, which are the closest to the chosen bead, are on the top of the list). Then, starting from the top of the list, we pick a j-th bead, and it can form a bond with the i-th bead with probability of bond formation, which is set constant in the simulation (if the bead on the k-th position in the list does not form a bond, the bead on the k+1-st position is checked, etc, until the end of the list). Therefore, our algorithm of bond formation favors bond formation of the spatially closest beads to the chosen one. Formed bond is identical to the backbone bonds of the chain. Then program iterates over all i-th beads until the end of the chain. If the chosen i-th bead already forms a bond, it can break the bond with probability of breaking a bond $\beta$, which is set constant during the simulation as well.

\begin{figure}[htbp]
	\includegraphics[width=\linewidth,height=\textheight,keepaspectratio]{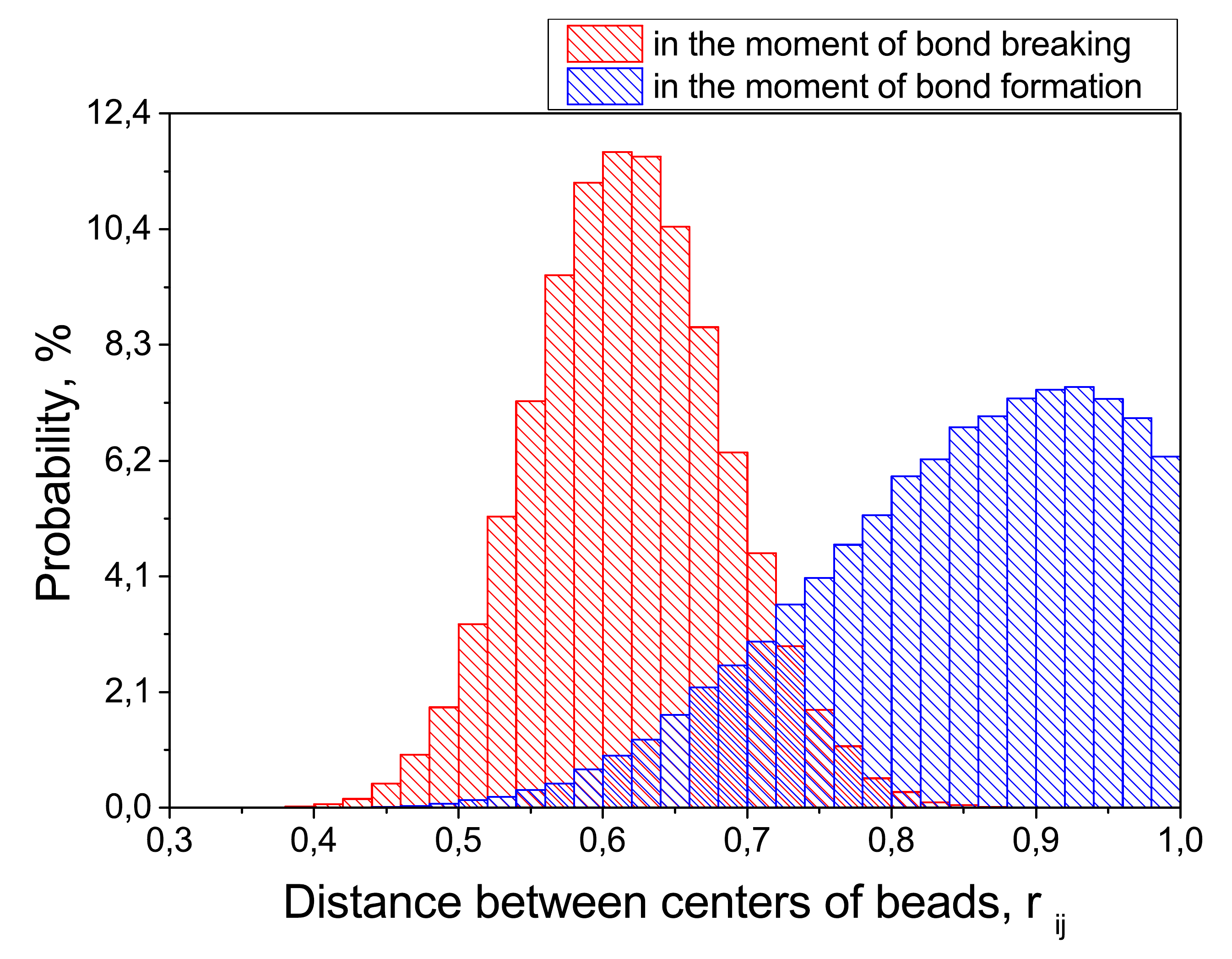}
    \caption{Probability distributions of distance between centers of beads in the moment of bond breaking (red) and bond formation (blue). Samples consist of 96514 distances for each histogram, data is normalized.}
    \label{FB_histograms}
\end{figure}

To study non-equilibrium properties of the chain, we needed to forbid self-intersections of the chain. To do so, we set large bond stiffness constant $K=150.0$, repulsion parameters, and small equilibrium bond length, which was equal to $0.6$ in all our simulations. By choosing these parameters we guaranteed the absence of self-intersections \cite{nikunen2007reptational}, but broke detailed balance (Figure \ref{FB_histograms}). Therefore, there is unequal probability distributions of finding two beads at a distance $r_{ij}$ at the moment of bond formation/breaking (Fig. \ref{FB_histograms}). This effect leads to energy gain and dissipation during bond formation/breaking process. However, dissipative and random forces acting between beads act as a thermostat in DPD \cite{groot1997dissipative}. Hence, if simulation parameters are chosen correctly, local gain and dissipation of energy after bond formation/breaking can be compensated by dissipative and random forces from surrounding beads. To ensure this compensation, we have chosen integration time step $\delta t = 0.02$ and noise level $\sigma =3$. If these parameters are chosen, the relaxation time of exponential decrease of temperature from $k_BT=10$ to $k_BT=1$ is around 10 DPD time steps \cite{groot1997dissipative}. In our study, minimal bond lifetime is $5\times10^4$ DPD time steps, so there is a strong evidence towards insufficiency of the effect of energy gain and dissipation during bond formation/breaking.

\subsection{Procedure of globule equilibration in SPS}
In our work, we studied conformational properties of equilibrium globule as a reference state. As we mentioned in the paper, we obtained stable transient state, similar to crumpled globule, during both types of coil-globule transition: in a poor solvent and induced by pairwise reversible bonds. To obtain equilibrium globule, we used crumpled globule state, formed in a weak poor solvent ($\chi\approx1.5$, SPS), after $t = 8\times10^7$ DPD time steps, as initial structure. The crumpled globule state is stabilized by topological interactions, therefore, we needed to facilitate self-intersections of the chain to obtain Gaussian globule. To do so, we set monomer-monomer and solvent-solvent repulsion parameters $a_{ii}$ $(i={1,2})$ equal to $25.0$, and monomer-solvent repulsion was equal to $a_{12} = a_{21} = 29.5$. Since $\chi$ value is determined from difference between $a_{ij}$ and $a_{ii}$ (if $a_{ii} = a_{jj}$ and, $a_{ij} = a_{ji}$) \cite{groot1997dissipative}, we conserved the solvent quality, but significantly reduced overall repulsion between polymer beads, therefore, facilitating self-intersections of the chain. We also reduced bond stiffness parameter from $K=150.0$ to $K=4.0$, and increased integration time step to $\delta t=0.04$. Therefore, as previously described in Ref. \cite{tamm2015anomalous}, we allowed chain to self-intersect, and conserved temperature $k_BT=1$ in the NVT-ensemble. We performed simulation with these parameters for $t = 5\times10^7$ DPD time steps. Then we changed simulation parameters to the initial values, described in Methods for SPS, and performed simulations for $t = 2\times10^7$ DPD time steps.

\subsection{Additional plots}
In this section, we provide additional dependencies, supporting the information in the paper.

\begin{figure*}[h!]
    \begin{subfigure}{0.45\textwidth}
	\includegraphics[width=\linewidth,height=\textheight,keepaspectratio]{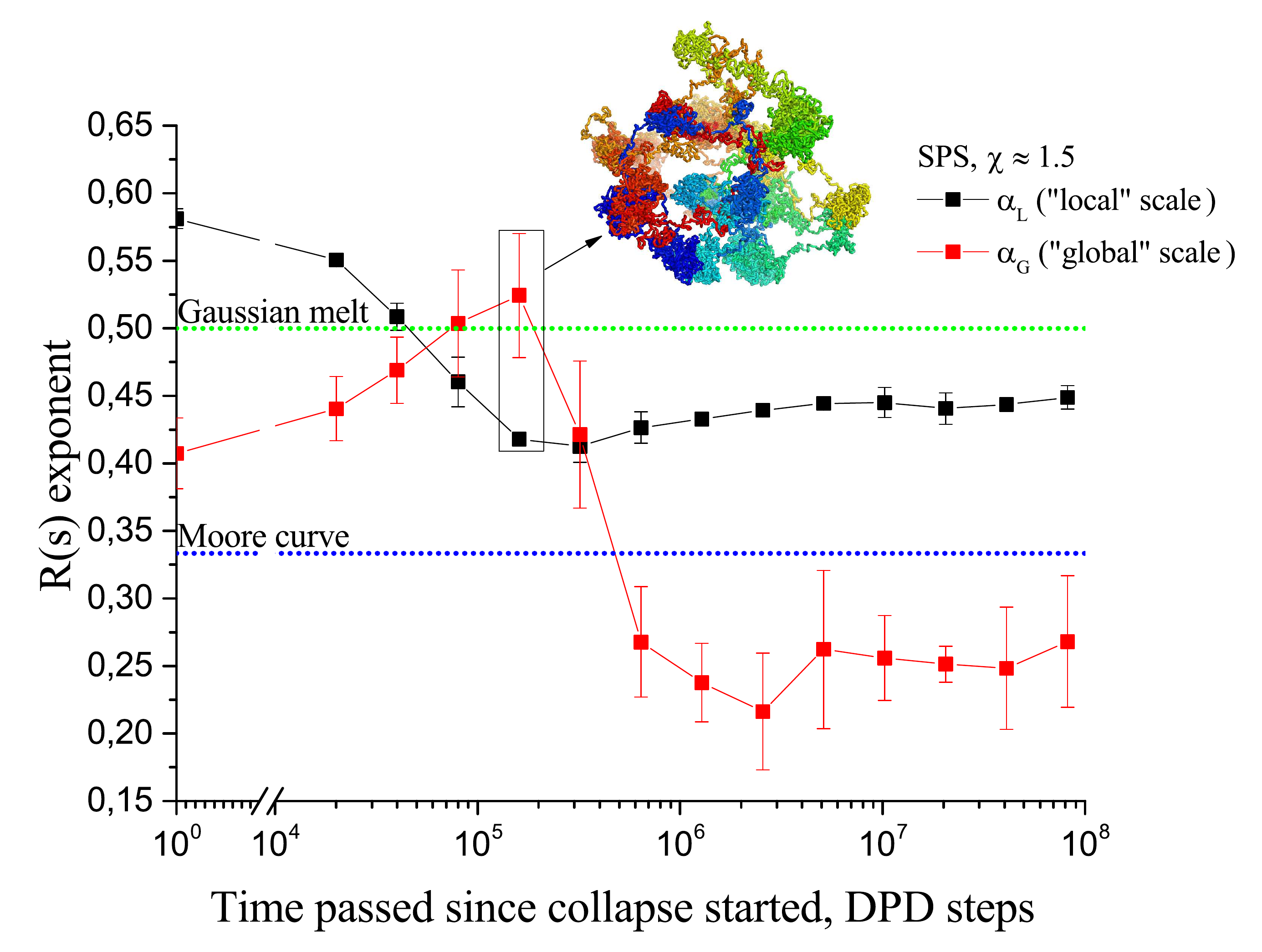}
	\caption{}
	\end{subfigure}
	\begin{subfigure}{0.45\textwidth}
	\includegraphics[width=\linewidth,height=\textheight,keepaspectratio]{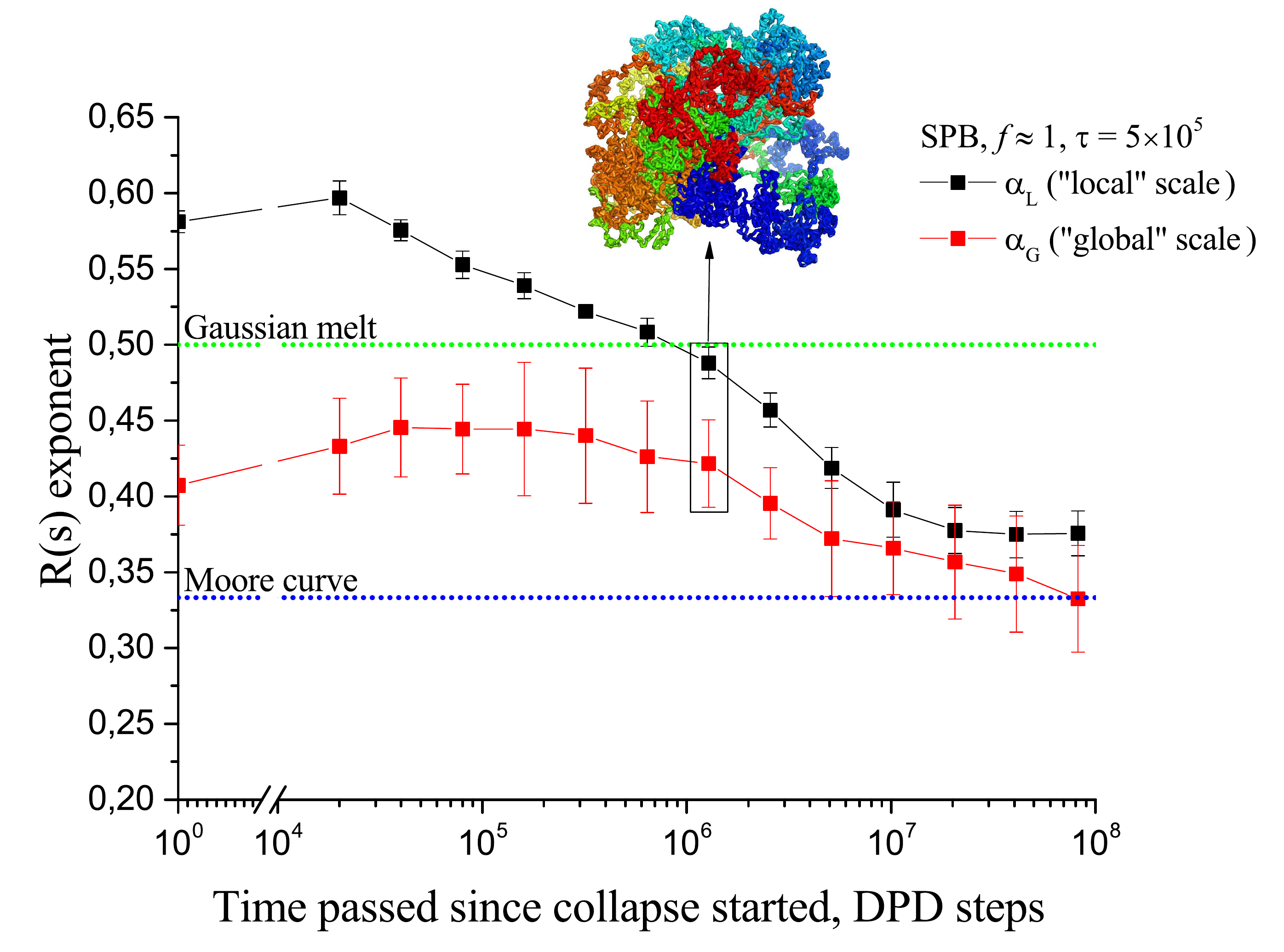}
	\caption{}
	\end{subfigure}
    \begin{subfigure}{0.45\textwidth}
	\includegraphics[width=\linewidth,height=\textheight,keepaspectratio]{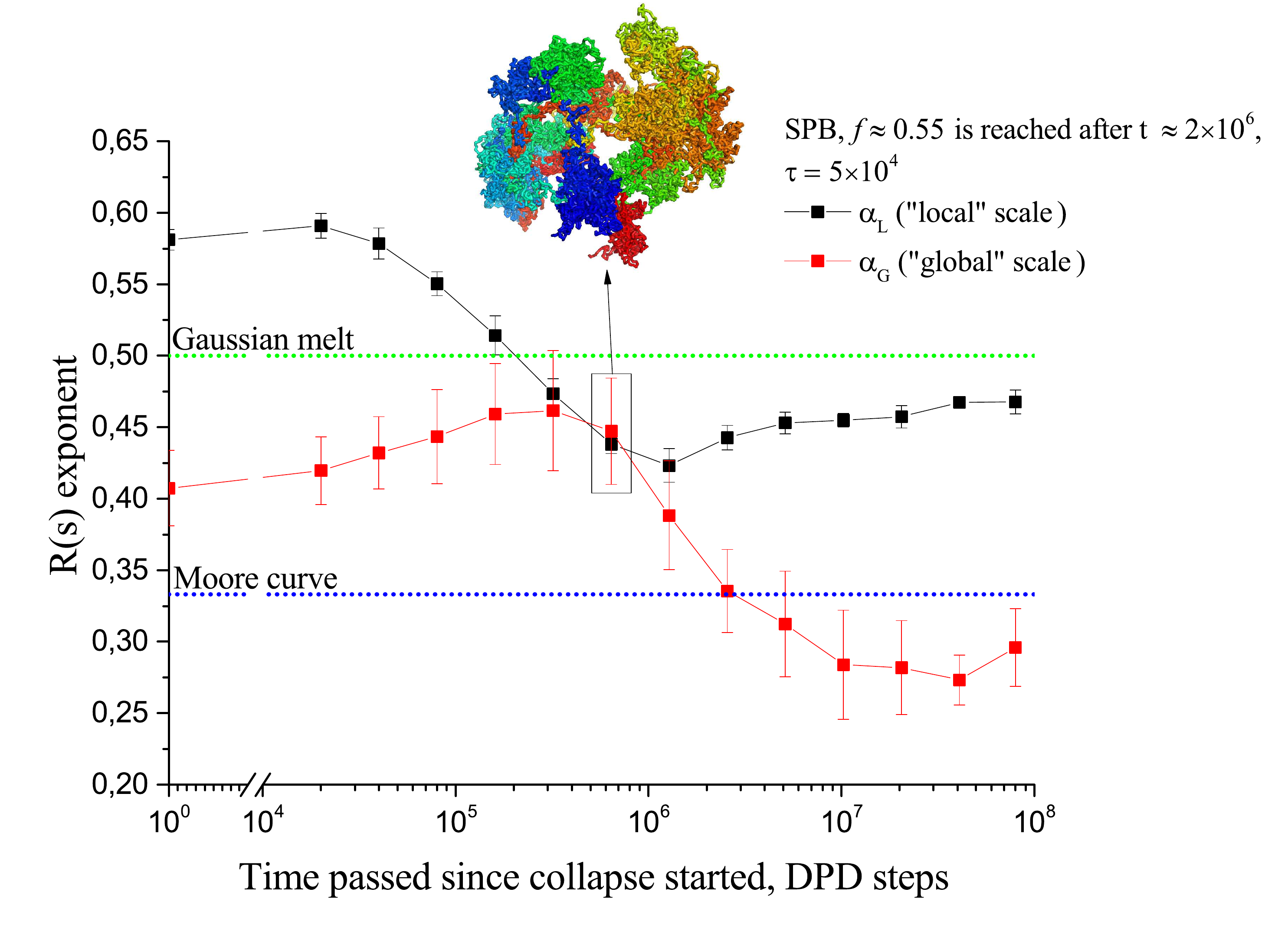}
	\caption{}
	\end{subfigure}
	\caption {$\alpha_L$ and $\alpha_G$ values (exponents of $R(s)$ dependencies on the "local" and "global" scale shown in black and red, respectively) for System in a Poor Solvent (SPS) in a weak poor solvent ($\chi\approx1.5$, Figure a), for System with Pairwise Bonds (SPB) with maximal fraction of bonds $f\approx 1$, bond lifetime is $\tau = 5\times10^5$ steps (Figure b), and for SPB with fraction of bonds $f\approx 0.55$ (Figure c), depending on the time passed since simulation started. It should be taken into account, however, that fraction of bonds $f\approx 0.55$ is reached only after $t=2\times10^6$ DPD time steps (Figure c), on the shorter time scale fraction of bonds was constantly increasing (Figure \ref{fracofbonds}b).}
	\label{slopesadditional}
\end{figure*}

\begin{figure}[h!]
	\includegraphics[width=\linewidth,height=\textheight,keepaspectratio]{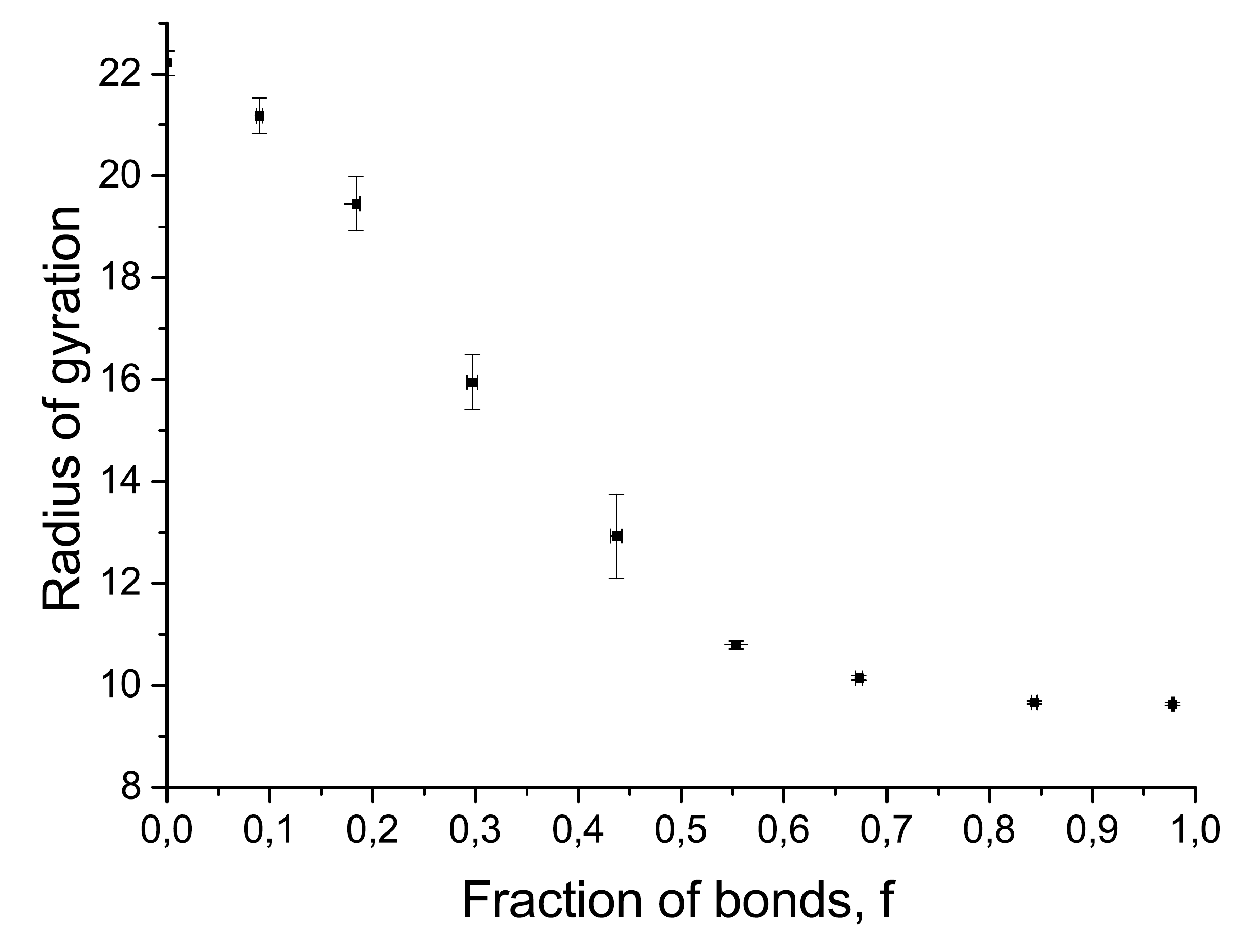}
    \caption{Dependency of the radius of gyration of the chain ($N=2\times10^4$) on the fraction of bonds with lifetime equal to $\tau = 5\times10^4$. Coil-globule transition occurs at $f \approx 0.3$. Compact conformations are obtained if $f\approx0.5$ and larger.}
    \label{coilglob}
\end{figure}

\begin{figure*}[h!]
    \begin{subfigure}{0.55\textwidth}
	\includegraphics[width=\linewidth,height=\textheight,keepaspectratio]{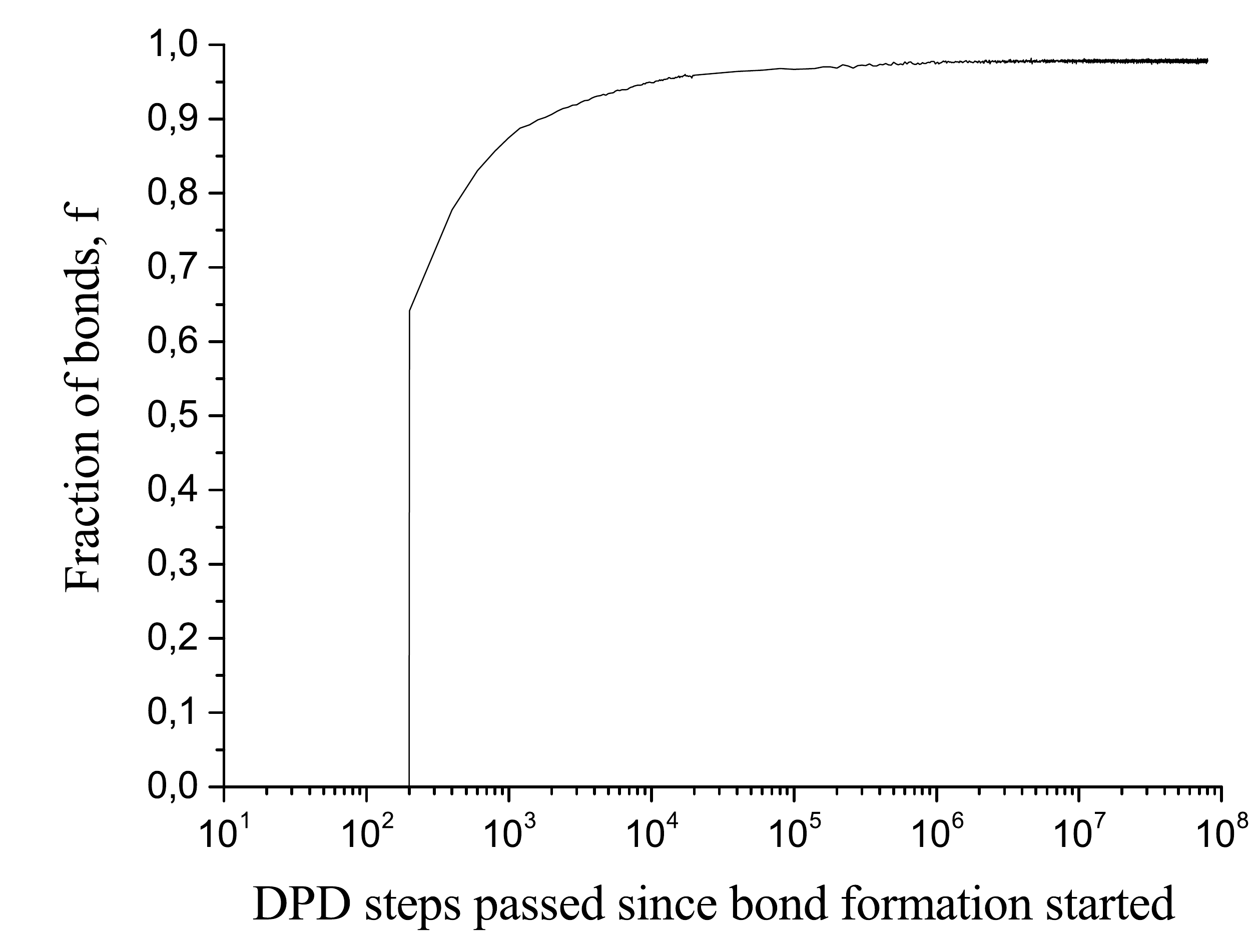}
	\caption{}
	\end{subfigure}
	\begin{subfigure}{0.55\textwidth}
	\includegraphics[width=\linewidth,height=\textheight,keepaspectratio]{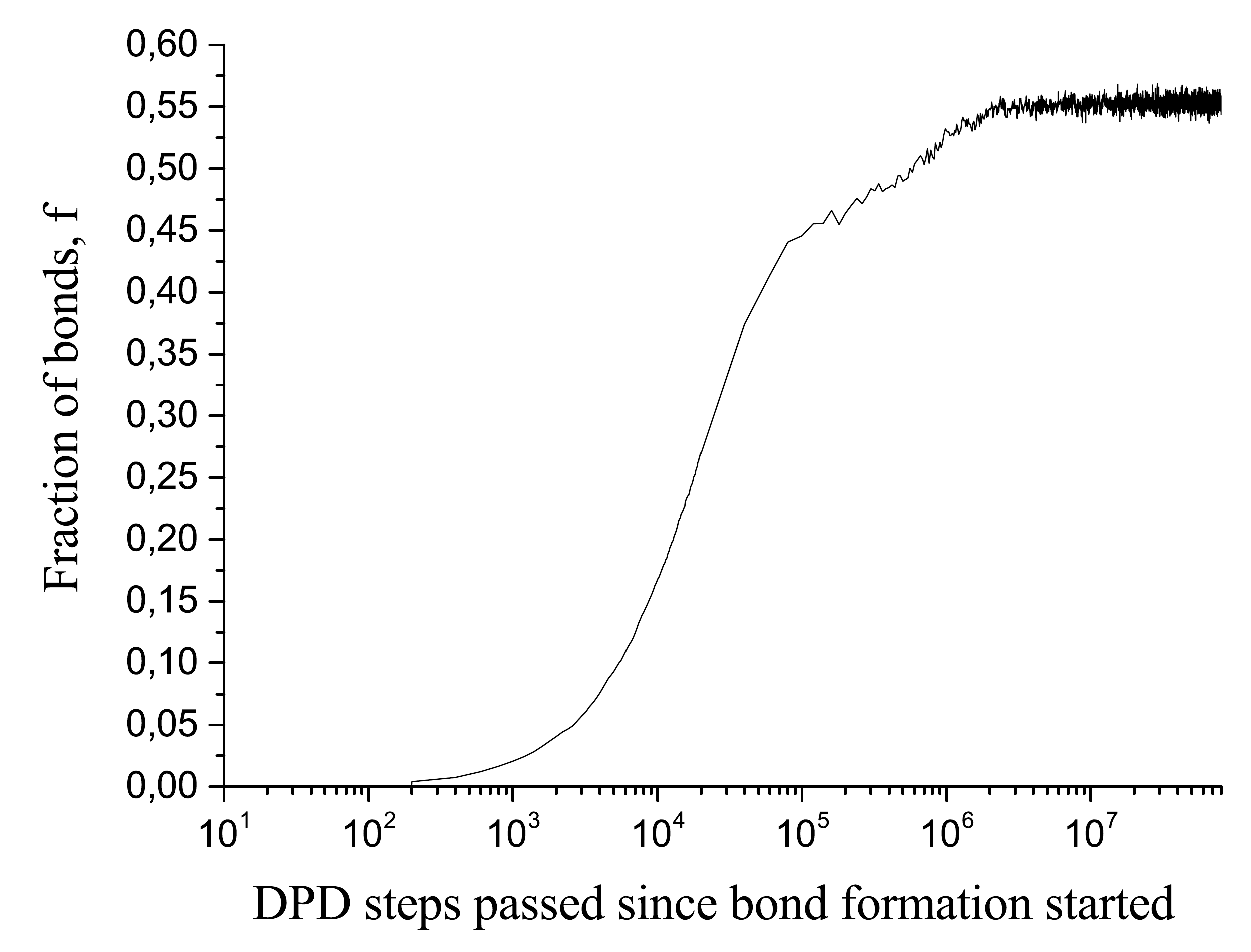}
	\caption{}
	\end{subfigure}
	\caption {Time dependencies of fraction of bonds in the SPB, bond lifetime is equal to $\tau = 5\times10^4$ steps. Bond formation started after $N_{stp}=200$ DPD time steps passed in every simulation. a) Probability of reversible bond formation is 1.0. As a result, fraction of bonds quickly stabilizes at value $f\approx 1$. b) Probability of reversible bond formation is $0.001$, therefore, fraction of bonds reaches the stable value $f\approx0.55$ relatively slowly.}
	\label{fracofbonds}
\end{figure*}

\begin{figure*}[h!]
	\includegraphics[width=\linewidth,height=\textheight,keepaspectratio]{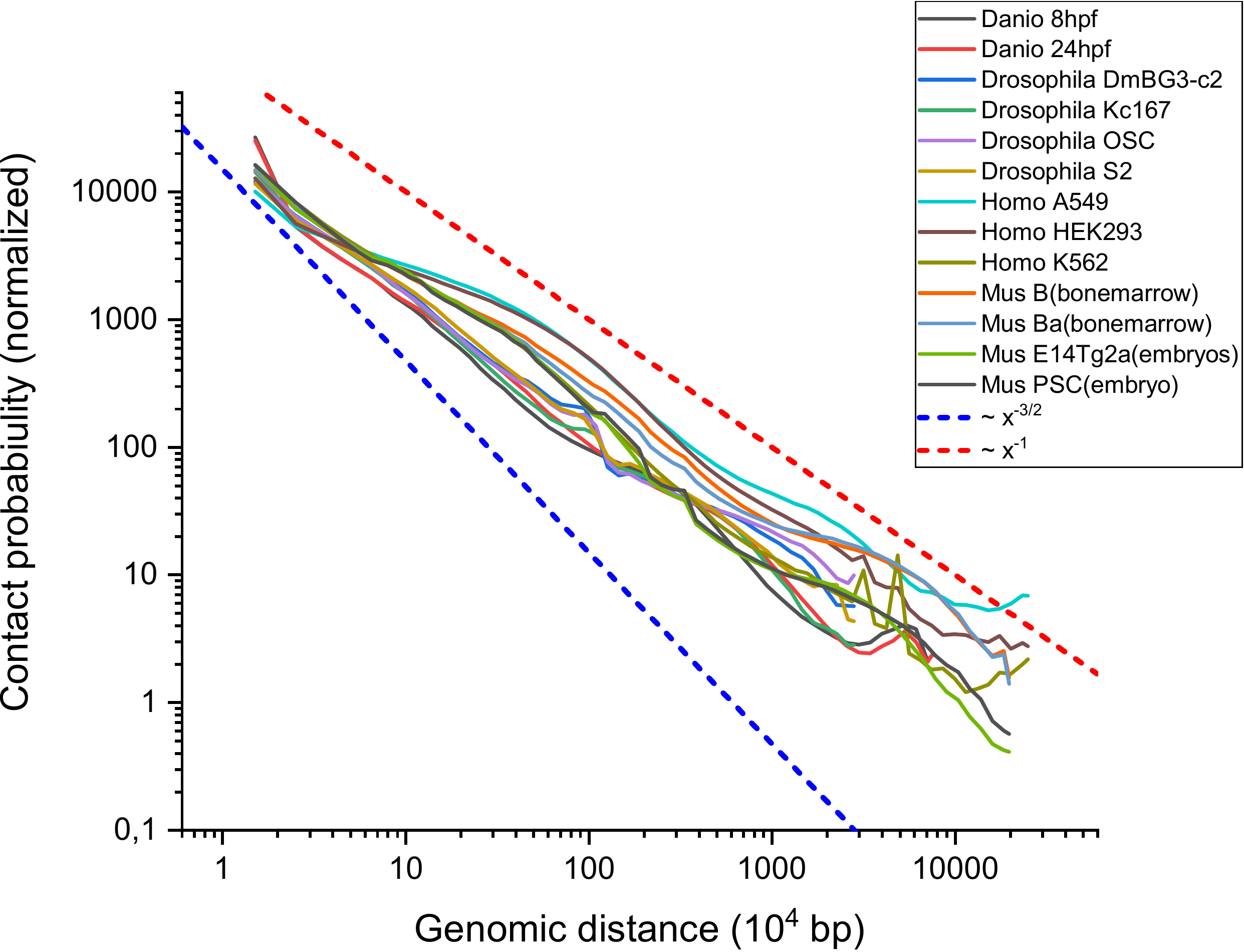}
	\caption{The figure shows the experimental contact probability dependencies for different species and cell lines. Blue dashed line corresponds to the equilibrium Gaussian globule and red dashed line corresponds to the fractal globule. Well-known dependence of the contact probability on genomic distance $P(s) \propto s^{-1}$ for Homo sapiens actually is not strictly enforced. For example, contact probability depends on considering scales and it works for another biological species as well. Moreover, scalings for different cell lines of the same species could significantly differ from each other on the same scale. Apparently minor protocol changes in Hi-C experiment, different equipment and different data processing have a significant impact on the final results. To plot this figure we used publicly available Hi-C datasets re-analyzed with distiller software with standard parameters (\textbf{https://github.com/mirnylab/distiller-nf}). 
		All replicates, if available, were merged together.
		The data for Homo sapiens was retrieved for: K562 cell line (GEO ID GSE63525, \textbf{https://www.ncbi.nlm.nih.gov/pubmed/25497547}), A549 GSE105600, \textbf{https://www.ncbi.nlm.nih.gov/pubmed/22955616},	HEpG2 GSE105381, \textbf{https://www.ncbi.nlm.nih.gov/pubmed/22955616};	Drosophila melanogaster cell lines Kc167, Dm3, OSC, S8: GSE69013 \textbf{https://www.ncbi.nlm.nih.gov/pubmed/26518482}; Danio rerio embryos GSE105013 \textbf{https://www.ncbi.nlm.nih.gov/pubmed/29972771};	Mus musculus data from GSE96611 \textbf{https://www.ncbi.nlm.nih.gov/pubmed/29335546} and from GSE96107 \textbf{https://www.ncbi.nlm.nih.gov/pubmed/29053968}. This analysis was performed by Alexandra Galitsyna.}
	\label{fig:biology}
\end{figure*}

\end{document}